\documentclass[lettersize,journal]{IEEEtran}
\ifCLASSINFOpdf

\else

\fi
\usepackage[cmex10]{amsmath}
\usepackage{amssymb}
\usepackage{cite}
\usepackage{graphicx}
\usepackage{array,color}
\usepackage{amsmath}
\usepackage{stfloats}
\usepackage{graphicx}
\usepackage{subfigure}
\usepackage{tabularx}
\usepackage{epsfig,epsf,color,balance,cite}
\usepackage{algorithmic}
\usepackage{algorithm} 
\usepackage{bm}
\usepackage{textcomp}
\usepackage{enumerate}
\usepackage{makecell}
\usepackage{supertabular}
\usepackage{tabularx}
\usepackage{array}
\usepackage{caption}
\allowdisplaybreaks[4]
\begin{document}

\title{ Beamforming Design for Double-Active-RIS-aided Communication Systems with Inter-Excitation}
	\author{Boshi Wang, Cunhua Pan\IEEEmembership{, Senior Member, IEEE}, Hong Ren\IEEEmembership{, Member, IEEE}, Zhiyuan Yu, Yang Zhang, \\Mengyu Liu, Gui Zhou\IEEEmembership{, Member, IEEE}
	\thanks{B. Wang, C. Pan, H. Ren, Z. Yu, Y. Zhang and M. Liu  are with National Mobile Communications Research Laboratory, Southeast University, Nanjing, China. (e-mail:\{boshi\_wang, cpan, hren, zyyu, y\_zhang, mengyuliu\}@seu.edu.cn). G. Zhou is with the Institute for Digital Communications, Friedrich-Alexander-University Erlangen-N{\"u}rnberg (FAU), 91054 Erlangen, Germany (e-mail: gui.zhou@fau.de). 
		
		\emph{Corresponding author: Cunhua Pan.}}
	
}

\maketitle
\vspace{-1.9cm}
\begin{abstract}
In this paper, we investigate a double-active-reconfigurable intelligent surface (RIS)-aided downlink wireless communication system, where a multi-antenna base station (BS) serves multiple single-antenna users with both double reflection and single reflection links.
Due to the signal amplification capability of active RISs, they can effectively mitigate the multiplicative fading effect. However, this also induces signal bouncing between the two active RISs that cannot be ignored. This phenomenon is termed as the “inter-excitation” effect and is characterized in the received signal by proposing a feedback-type model. 
Based on the signal model, we formulate a weighted sum rate (WSR) maximization problem by jointly optimizing the beamforming matrix at the BS and the reflecting coefficient matrices at the two active RISs, subject to power constraints at the BS and active RISs, as well as the maximum amplification gain constraints of the active RISs.
To solve this non-convex problem, we first transform the problem into a more tractable form using the fractional programming (FP) method.
Then, by introducing auxiliary variables, the problem can be converted into an equivalent form that can be solved by using a penalty dual decomposition (PDD) algorithm. 
Finally, simulation results indicate that it proposed scheme outperforms benchmark schemes with single active RIS and double passive RISs in terms of achievable rate. 
Furthermore, the results demonstrate that the proposed scheme can enhance the WSR by 30\% compared to scenarios that do not take this effect into account when the maximum amplification gain is 40 dB. 
Additionally, the proposed scheme is capable of achieving high WSR performance at most locations where double active RISs are deployed between the BS and the users, thereby providing greater flexibility in their positioning.
\end{abstract}
\begin{IEEEkeywords}
Reconfigurable intelligent surface (RIS), active RIS, double RISs, inter-RIS signal reflection.
\end{IEEEkeywords}

\IEEEpeerreviewmaketitle
\section{Introduction}
\IEEEPARstart{R}{econfigurable} intelligent surfaces (RISs) are  technologies that significantly improve wireless communication performance by proactively controlling the wireless propagation environment. Besides,
due to the benefits of low cost, low power consumption, programmability, and easy deployment, RIS has become one of the key technologies of 6G, attracting wide research attention from both academia and industry\cite{8910627,9475160,9847080}.
Recently, there have been a lot of efforts devoted to the transmission design of RIS-aided wireless communication systems, including single-user systems \cite{8647620}, multi-user systems \cite{8811733}, multi-cell systems \cite{9090356}, integrated sensing and communication (ISAC) systems \cite{9913311,9729741}, mobile edge computing \cite{9133107}, simultaneous wireless
information and power transfer (SWIPT) \cite{9110849} and non-orthogonal multiple access (NOMA) \cite{NOMA}.

Most of the above-mentioned contributions mainly considered one or multiple distributed RISs, where each RIS independently reflects signals to users without the cooperation among multiple RISs.
However, it is limited for deploying single RIS for one wireless link, which may not fully exploit the potential of RIS in enhancing communication performance. 
This is because the maximum size of RIS is constrained by practical limitations, and a single RIS may only provide limited passive beamforming gain for each user. Additionally, single reflecting path provided by a single RIS may not be able to bypass the complex environments \cite{9771079}. Moreover, the single RIS or the  non-cooperative multiple distributed RISs can only achieve a power scaling order proportional to the square of the number of reflecting elements.
In order to overcome the above-mentioned limitations, the authors of \cite{9060923} studied a cooperative double-RIS-aided single-user wireless communication system.
In \cite{9060923}, the authors showed that the beamforming gain of the double passive RIS increases with the fourth power of the total number of reflecting elements of the RIS, significantly higher than the quadratic growth of the beamforming gain of the single passive RIS with the same number of total reflecting elements. 
The authors of \cite{9362274,9363566} extended the scenario to include multiple users and demonstrated that the deployment of cooperative double passive RISs can improve communication performance over the single passive RIS. Subsequently, the authors of \cite{9829192} extended the  double-RIS-aided systems to a more general cooperative multi-RIS-aided system, indicating that it is more advantageous to distribute the reflecting elements into cooperative multiple RISs than deploying them on a single RIS.


However, the high beamforming gain achieved by double-RIS-aided systems comes at the price of increased multiplicative fading. Specifically, the equivalent pathloss of the reception signal is the product of the individual channel pathloss. This limitation is acceptable only when one of the RIS is deployed either near the base station (BS) or  users, which limits the wider application of double-RIS-aided systems.
As a promising solution to this issue, the authors of \cite{9998527,9377648} proposed the novel concept of the active RISs. With additionally equipped reflection-type amplifiers, active RISs can not only adjust the phase shift, but change the amplitude of the incident signal as well.
Recent studies have extensively explored the potential benefits of active RIS in various scenarios, including ISAC systems\cite{10319318,2024arXiv240204532L}, SWIPT systems\cite{10012424,9810984}, physical layer security\cite{2024arXiv240202122Z}, etc.
Inspired by the substantial success of active RIS, the authors of \cite{10571224,10190735,10178078} incorporated active RIS into a double-RIS-aided system. Their findings indicate that the data rate of the proposed system significantly surpasses that of the double-passive-RIS-aided system with a small number of reflecting elements.

In multi-RIS-aided systems, the signal is reflected back and forth between the RISs, typically with negligible power due to the significant multiplicative fading effect in passive RISs. However, since active RISs are equipped with amplifiers, the interaction between multiple RISs cannot be ignored. As this reciprocal process stabilizes, the reflected signal at the active RIS includes not only the original signal and thermal noise but also the previously reflected back signal. This phenomenon is referred to as the “inter-excitation” effect.
Due to the nanosecond-level processing delay of the active RIS and the microsecond-level propagation delay of electromagnetic wave, the incident and reflected signals of RISs carry the same symbols in one time slot \cite{9998527}. 
Therefore, the re-incident signals to the active RIS owing to inter-excitation may carry useful information and should not be treated simply as interference. 
However, it remains unknown what the potential impact of inter-excitation is and how to perform beamforming design considering inter-excitation.

Inspired by the above-mentioned background, we investigate the beamforming design for a double-active-RIS-aided multi-user multi-input single-output (MU-MISO) system with both double and single reflecting links, while considering the inter-excitation effect between the active RISs in the system.
The main contributions of this paper are summarized as follows:

\begin{itemize}
	\item[1)] To the best of our knowledge, this is the first work that investigates the impact of the inter-excitation effect between the double active RISs.
	A feedback signal model is developed to characterize the inter-excitation effect between two active RISs, and the expression for the reflected signals is derived in the steady state of inter-excitation.
	Based on the inter-excitation model, we formulate a weighted sum rate (WSR) maximization problem for the double-active-RIS-aided MU-MISO system subject to the power constraints at the BS and active RISs, as well as the maximum amplification gain constraints of the active RISs.

	\item[2)] We propose a algorithm based on the double-loop penalty dual decomposition (PDD) framework to jointly optimize the beamforming matrix of the BS and the reflecting coefficient matrices of the active RISs.
	Specifically, the objective function is first transformed into a more compact form by using the fractional programming (FP) method. Then, we introduce the auxiliary variables and equality constraints to tackle the inverse form of inter-excitation matrices, which are addressed using PDD framework. 
	Subsequently,  alternating optimization (AO) approach is adopted to decouple the problem into subproblems, which are solved by the ellipsoid method and bisection search method.
	\item[3)] 
	Finally, simulation results indicate that significant data rate improvement can be achieved in double active RIS-aided systems than single active RIS-aided systems and double passive RIS-aided systems. 
	Furthermore, it is crucial to take into account the inter-excitation effect, particularly when the distance between the two active RISs is not relatively large and the maximum amplification gain is considerable.
	The inter-excitation exhibits the potential to enhance the WSR performance with limited maximum amplification gain of the active RISs' amplifiers.
\end{itemize}

The remainder of this paper is organized as follows. In
Section II, we introduce  the system model and problem formulation. The algorithm is developed in Section III.
Simulation results and conclusion are reported in Sections IV and V, respectively. 

\textbf{Notations:} The mathematical notations used in this paper are as follows. Operations $\left | a \right |$, $a^*$, $\textrm{Re}\left \{a \right \}$ and $\angle a$ denote the absolute value, conjugate, real part and angle of a complex number $a$, respectively. 
$\left [ a \right ] ^d_c$ denotes that the mapping of $a$ to the interval $\left [c,d  \right ]$, i.e.,  $\left [ a \right ] ^d_c$ is equal to the value closest to $a$ in the interval $\left [c,d  \right ]$.
 $\textrm{sgn}(a)$ denotes the sign function, which is defined such that it equals 1 if $a$ is greater than 0, it equals 0 if $a$ is equal to 0, and it equals -1 if $a$ is less than 0.
Vectors and matrices are denoted by boldface lowercase letters and boldface uppercase letters, respectively. $\left \|\mathbf{x}\right \|_2$ denotes the $l_2$-norm of the vector $\mathbf{x}$.
$\left \|\mathbf{x}\right \|_{\infty}$ denotes the infinity norm of the vector $\mathbf{x}$. $\left \|\mathbf{X}\right \|_F$ denotes the Frobenius norm of matrix $\mathbf{X}$. $\textrm{Diag}(x_1,\cdots,x_N)$ denotes an $N \times N$ diagonal matrix, where the complex numbers $x_1,\cdots,x_N$ are the diagonal elements. 
$\left [ \mathbf{X} \right ] _{i,j}$ denotes the element in the 
$i$-th row and $j$-th column of matrix $\mathbf{X}$.
$\textrm{diag}(\mathbf{X})$ denotes a vector consisting of the main diagonal elements of the matrix $\mathbf{X}$. $\mathbf{X}\odot\mathbf{Y}$ denotes the Hadamard product between the matrix $\mathbf{X}$ and $\mathbf{Y}$. $\mathbf{X}\otimes \mathbf{Y}$ denotes the Kronecker product between the matrix $\mathbf{X}$ and $\mathbf{Y}$. $(\cdot )^T$, $(\cdot )^H$ and $\textrm{Tr}(\cdot )$ denote the transpose operation, the conjugate transpose operation, the trace operation, respectively. $\mathbb{C}$ denotes the space of complex numbers. $\mathbb{R}$ denotes the space of real numbers. $\mathbb{S}^{n}_{++}$ denotes the set of symmetric positive definite $n\times n$ matrices. $\mathcal{CN}\left ( \boldsymbol{\mu} , \boldsymbol{\Sigma} \right )$ represents the distribution of a circularly symmetric complex Gaussian random vector with mean vector $\boldsymbol{\mu}$ and covariance matrix $\boldsymbol{\Sigma}$. $j\triangleq\sqrt{-1}$ denotes the imaginary unit. $\mathcal{O}(\cdot )$ denotes the standard big-O notation.

\section{System Model and Problem Formulation}
\subsection{System  Model}
\begin{figure}
	\centering
	\includegraphics[width=3.5in]{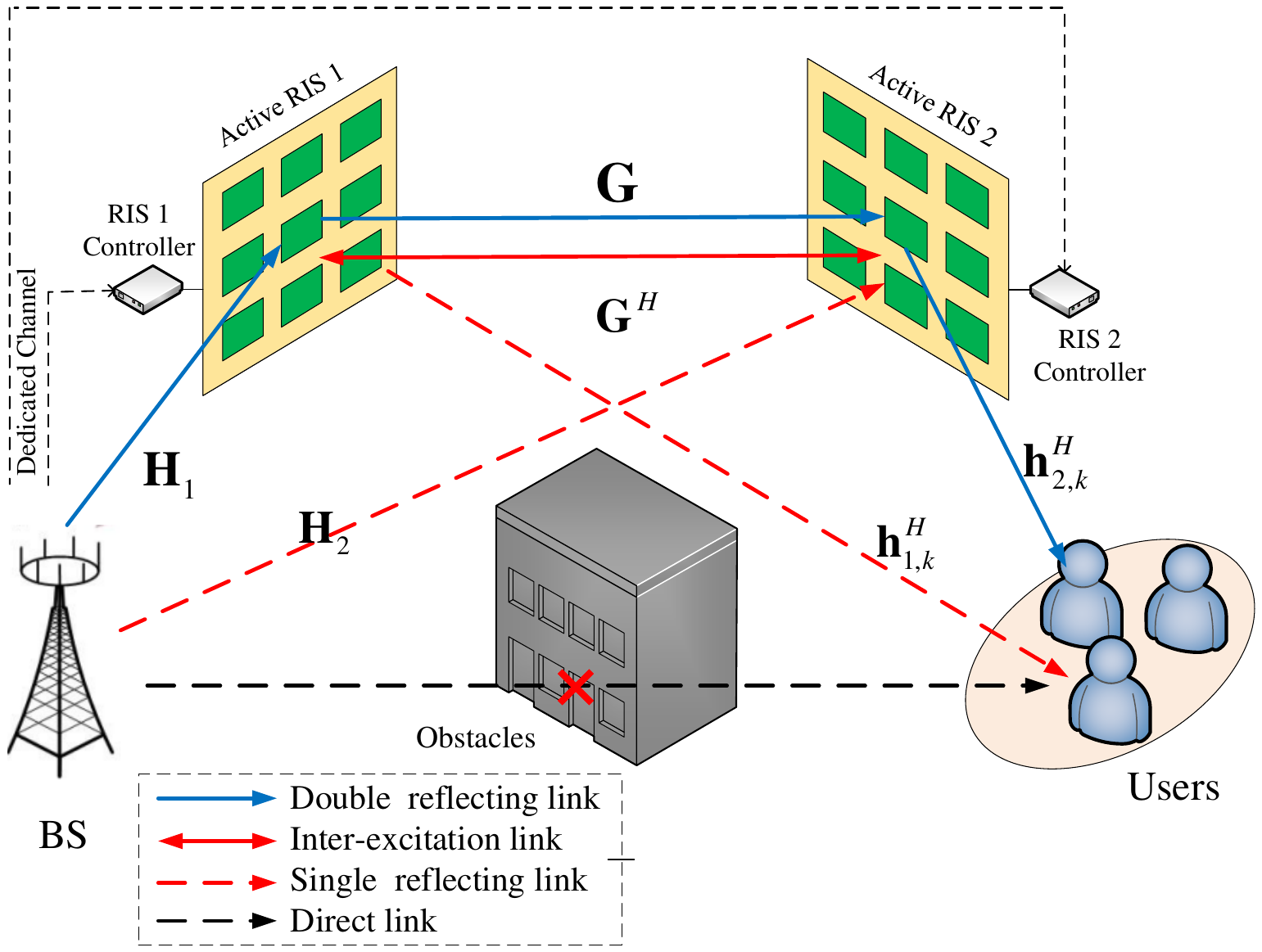}
	\captionsetup[figure]{justification=centering}
	\caption{Double-active-RIS-aided system with inter-excitation effect.}
	\label{System_model}
\end{figure}

We consider a cooperative double-active-RIS-aided MU-MISO communication system shown in Fig. \ref{System_model}, where double active RISs namely RIS 1 and RIS 2 are deployed to assist the downlink transmission from an $N$-antenna BS to $K$ single-antenna users indexed by $\mathcal{K}=\left \{1,\cdots,K \right \}$. It is assumed that the direct links from the BS to the users are severely blocked by obstacles and can be ignored. Double active RISs can establish communication links between the BS and the users, as shown in Fig. \ref{System_model}. Active RIS 1 and RIS 2 are equipped with $M_1$ and $M_2$ reflecting elements, leading to a total of $M=M_1 + M_2$ reflecting elements. For convenience, let us define the sets of RISs and elements of each RIS as $\mathcal{L}=\left \{1, 2 \right \}$ and $\mathcal{M}_l=\left \{1, \cdots, M_l \right \}, \forall l\in \mathcal{L}$, respectively.
\subsection{Transmit Signal Model}
The transmit signal from the BS to the users is expressed as
\begin{equation}\label{transmitsignal}
	{{\mathbf{x}}} ={\mathbf{W}}\mathbf{s} = \sum_{k=1}^{K} {\mathbf{w}_k}s_k,
\end{equation}
where $\mathbf{W}=\left [{\mathbf{w}_1}, \cdots,{\mathbf{w}_K}  \right ] \in \mathbb{C}^{N\times K}$ denotes the beamforming matrix and $\mathbf{s}=\left [{s_1}, \cdots,{s_K}  \right ]^T\in \mathbb{C}^K$ is the symbol vector transmitted to the $K$ users following the distribution of independent complex Gaussian, i.e., $\mathbf{s}\sim \mathcal{CN}\left ( 0,\mathbf{I}_{K} \right )$. Thus, the BS transmit power constraint  can be expressed as 
\begin{equation}
	\mathbb{E} \{ \left \| \mathbf{W}\mathbf{s} \right \|^2_2    \} =\left \| \mathbf{W} \right \|^2_F=\sum_{k=1}^{K} \left \| {\mathbf{w}_k}  \right \|_2 ^2 \leq P_{\textrm{BS}}^{\textrm{max}},
\end{equation}
where $P_{\textrm{BS}}^{\textrm{max}}$ is the maximum transmit power of the BS.
\subsection{Channel Model}
Let ${\mathbf{H}_1}\in\mathbb{C}^{M_{1}\times N}$, ${\mathbf{H}_2}\in \mathbb{C}^{M_{2}\times N}$, ${\mathbf{G}}\in\mathbb{C}^{M_{2}\times M_{1}}$, ${\mathbf{h}_{1,k}}\in\mathbb{C}^{M_{1}\times 1}$ and ${\mathbf{h}_{2,k}}\in\mathbb{C}^{M_{2}\times 1}$ denote the  channels from the $\textrm{BS}$ to $\textrm{RIS}$ 1, from the $\textrm{BS}$ to $\textrm{RIS}$ 2, from $\textrm{RIS}$ 1 to $\textrm{RIS}$ 2, from $\textrm{RIS}$ 1 to the $k$-th user, from $\textrm{RIS}$ 2 to the $k$-th user, respectively, with $k\in\mathcal{K}$. 
It is assumed that all the perfect channel state information (CSI) is known at the BS. 
\subsection{Active RIS Model with Inter-excitation Effect}
As shown in Fig. 1, for ease of distinction, we designate the RIS near the BS as RIS 1 and the RIS near the users as RIS 2.
For each active RIS $l$ with $l\in \mathcal{L}$,  the reflection coefficient matrix is denoted as $\boldsymbol{\Psi }_{l} = \textrm{Diag}( \psi_{l,1},\cdots, \psi_{l,M_l})$, where $\psi_{l,m_l}=a_{l,m_l}e^{j\theta_{l,m_l}}$. Herein, $1\le a_{l,m_l}\le a_{\textrm{max}}$ and $\theta_{l,m_l}\in\left [ 0,2\pi \right ] $ are the amplitude and the phase shift of each element, where $a_{\textrm{max}}$ is the maximum amplitude gain of the amplifier.  


As the system employs two active RISs, the mutual influence between them should be taken into account, i.e. the signal bouncing between two active RISs in the double reflection link.
Specifically, considering the downlink transmission, the signal can be reflected by RIS 1 to RIS 2, and RIS 2 can also reflect the signal back to RIS 1.
The signal is reflected back and forth between the two RISs until the reflected signal reaches a stable state, which is termed as the inter-excitation effect.
In conventional cooperative multi-passive-RIS-aided system, the limited power of the reflected signal results in significant attenuation of secondary reflecting signals, allowing them to be ignored.
However, due to the amplification function of the active RIS, the energy of the reflected signal is enhanced, rendering the inter-excitation effect significant and unavoidable.

Taking into account the thermal noise at the active RISs, which is significant due to the presence of reflection-type amplifiers, the instantaneous reflected signals at the active RIS 1 and active RIS 2 are respectively expressed as follows
\begin{subequations}\label{instantaneous_y}
 \begin{align} 
 	{\tilde{{\mathbf{y}}}_{1}}&=\boldsymbol{\Psi }_{1} (\mathbf{H}_1{{\mathbf{x}}}+{\mathbf{G}^H}{\tilde{{\mathbf{y}}}_{2}}+\mathbf{n}_{1})\label{y_1}, \\
	{\tilde{{\mathbf{y}}}_{2}} &=\boldsymbol{\Psi }_{2} (\mathbf{H}_2{{\mathbf{x}}}+{\mathbf{G}}{\tilde{{\mathbf{y}}}_{1}}+\mathbf{n}_{2}) \label{y_2},
\end{align}
\end{subequations}
where $\mathbf{n}_l\sim \mathcal{CN}\left ( 0,\sigma^{2}_{l}\mathbf{I}_{M_l} \right )$, $\forall l\in \mathcal{L}$, denotes the thermal noise at the active RIS $l$ with the noise power of $\sigma^{2}_{l}$. 
Due to the nanosecond-level processing delay associated with the active RIS\cite{9998527} and the microsecond-level propagation delay of electromagnetic waves, both $\tilde{\mathbf{y}}_1$ and $\tilde{\mathbf{y}}_2$ carry information of the same time slot. 
Therefore, the incident signal from another RIS cannot be simply eliminated as interference in the beamforming design, but rather contains the useful signal.

Hence, by substituting (\ref{y_1}) into (\ref{y_2}) and (\ref{y_2}) into (\ref{y_1}), respectively, we have:
\begin{subequations}
\begin{align} 
 {\tilde{{\mathbf{y}}}_{1}}&=\underbrace {\boldsymbol{\Psi }_{1}(\mathbf{H}_1+{\mathbf{G}^H}\boldsymbol{\Psi }_{2}\mathbf{H}_2){{\mathbf{x}}}}_{\text {Original signal}}+\underbrace {\boldsymbol{\Psi }_{1}{\mathbf{G}^H}\boldsymbol{\Psi }_{2}{\mathbf{G}}{\tilde{{\mathbf{y}}}_{1}}}_{\text {Inter-excitation signal}}\nonumber\\&+\underbrace {\boldsymbol{\Psi }_{1}{\mathbf{G}^H}\boldsymbol{\Psi }_{2}{{\mathbf{n}_{2}}}+\boldsymbol{\Psi }_{1}\mathbf{n}_{1}}_{\text {Dynamic noise}},\label{y_1_fb}\\
  {{\tilde{\mathbf{y}}_{2}}} &=\underbrace {\boldsymbol{\Psi }_{2}(\mathbf{H}_2+{\mathbf{G}}\boldsymbol{\Psi }_{1}\mathbf{H}_1){{\mathbf{x}}}}_{\text {Original signal}}+\underbrace {\boldsymbol{\Psi }_{2}{\mathbf{G}}\boldsymbol{\Psi }_{1}{\mathbf{G}^H}{{\tilde{\mathbf{y}}_{2}}}}_{\text {Inter-excitation signal}}\nonumber\\&+\underbrace {\boldsymbol{\Psi }_{2}{\mathbf{G}}\boldsymbol{\Psi }_{1}{{\mathbf{n}_{1}}}+\boldsymbol{\Psi }_{2}\mathbf{n}_{2}}_{\text {Dynamic noise}}. \label{y_2_fb}
\end{align}
\end{subequations}

Note that  (\ref{y_1_fb}) and (\ref{y_2_fb}) can be interpreted as a feedback model. Subsequently, by transferring the instantaneous inter-excitation signals of the active RISs from equations (\ref{y_1_fb}) and (\ref{y_2_fb}) to the left side, we have:
\begin{subequations}
\begin{align} 
	(\mathbf{I}_{M_1}-\boldsymbol{\Psi }_{1}{\mathbf{G}^H}\boldsymbol{\Psi }_{2}{\mathbf{G}}){{\tilde{\mathbf{y}}_{1}}}&=\boldsymbol{\Psi }_{1}(\mathbf{H}_1+{\mathbf{G}^H}\boldsymbol{\Psi }_{2}\mathbf{H}_2){{\mathbf{x}}}\nonumber\\&+\boldsymbol{\Psi }_{1}{\mathbf{G}^H}\boldsymbol{\Psi }_{2}{{\mathbf{n}_{2}}}+\boldsymbol{\Psi }_{1}\mathbf{n}_{1} ,\label{y_1_fc}\\
	(\mathbf{I}_{M_2}-\boldsymbol{\Psi }_{2}{\mathbf{G}}\boldsymbol{\Psi }_{1}{\mathbf{G}^H}){{\tilde{\mathbf{y}}_{2}}} &=\boldsymbol{\Psi }_{2}(\mathbf{H}_2+{\mathbf{G}}\boldsymbol{\Psi }_{1}\mathbf{H}_1){{\mathbf{x}}}\nonumber\\&+\boldsymbol{\Psi }_{2}{\mathbf{G}}\boldsymbol{\Psi }_{1}{{\mathbf{n}_{1}}}+\boldsymbol{\Psi }_{2}\mathbf{n}_{2}.\label{y_2_fc}
\end{align}
\end{subequations}

It can be observed from   (\ref{y_1_fc}) and (\ref{y_2_fc}) that the system attains a steady state when none of the determinants of $(\mathbf{I}_{M_1}-\boldsymbol{\Psi }_{1}{\mathbf{G}^H}\boldsymbol{\Psi }_{2}{\mathbf{G}})$ and $(\mathbf{I}_{M_2}-\boldsymbol{\Psi }_{2}{\mathbf{G}}\boldsymbol{\Psi }_{1}{\mathbf{G}^H})$ are equal to zero\cite{9998527}. 
To approximate the estimated matrices $\boldsymbol{\Psi }_{1}{\mathbf{G}^H}\boldsymbol{\Psi }_{2}{\mathbf{G}}$ and $\boldsymbol{\Psi }_{2}{\mathbf{G}}\boldsymbol{\Psi }_{1}{\mathbf{G}^H}$ in terms of their order of magnitude, we assume that $\boldsymbol{\Psi }_{1}=a_{\textrm{max}}\mathbf{I}_{M_1}$, $\boldsymbol{\Psi }_{2}=a_{\textrm{max}}\mathbf{I}_{M_2}$ and $\mathbf{G}\sim \mathcal{CN}\left ( 0,\beta_r\mathbf{I}_{M_1}\otimes\mathbf{I}_{M_2} \right )$, where $\beta_r$ is the large-scale path loss of the double reflection link. Subsequently, $ \mathbb{E} \left \{ \boldsymbol{\Psi }_{1}{\mathbf{G}^H}\boldsymbol{\Psi }_{2}{\mathbf{G}} \right \}=a_{\textrm{max}}^2\beta_r\mathbf{I}_{M_1}$ and $ \mathbb{E} \left \{ \boldsymbol{\Psi }_{2}{\mathbf{G}}\boldsymbol{\Psi }_{1}{\mathbf{G}^H} \right \}=a_{\textrm{max}}^2\beta_r\mathbf{I}_{M_2}$.
Thus, the amplification gain provided by the RISs is insufficient to compensate for the passloss for the double reflection link when the distance between the two active RIS is considerable. 
For example, when the distance between the two active RIS is 20 meters, $\beta_r=-60\ \textrm{dB}$ with the operating frequency of 2.4 GHz, which is significantly larger than the typical maximum amplitude of the active RIS, i.e., $a_{\textrm{max}}^2=40 \ \textrm{dB}$\cite{9377648,10134546}.
Consequently, the maximum eigenvalue of the two matrices $\boldsymbol{\Psi }_{1}{\mathbf{G}^H}\boldsymbol{\Psi }_{2}{\mathbf{G}}$ and $\boldsymbol{\Psi }_{2}{\mathbf{G}}\boldsymbol{\Psi }_{1}{\mathbf{G}^H}$ are generally much smaller than 1 in the general case. This ensures non-zero determinants for both $(\mathbf{I}_{M_1}-\boldsymbol{\Psi }_{1}{\mathbf{G}^H}\boldsymbol{\Psi }_{2}{\mathbf{G}})$ and $(\mathbf{I}_{M_2}-\boldsymbol{\Psi }_{2}{\mathbf{G}}\boldsymbol{\Psi }_{1}{\mathbf{G}^H})$. Define $\mathbf{\Xi}_1=(\mathbf{I}_{M_1}-\boldsymbol{\Psi }_{1}{\mathbf{G}^H}\boldsymbol{\Psi }_{2}{\mathbf{G}})^{-1}$ and $\mathbf{\Xi}_2=(\mathbf{I}_{M_2}-\boldsymbol{\Psi }_{2}{\mathbf{G}}\boldsymbol{\Psi }_{1}{\mathbf{G}^H})^{-1}$ as the inter-excitation matrices of active RIS 1 and active RIS 2, respectively.

	Therefore, by multiplying both sides of equation (\ref{y_1_fc}) by $\mathbf{\Xi}_1$ and equation (\ref{y_2_fc}) by $\mathbf{\Xi}_2$, the stabilized reflected signals of active RIS 1 and RIS 2 can be respectively expressed as follows
\begin{subequations}\label{stabilized_y}
\begin{align} 
	{{\mathbf{y}_{1}}}&=\mathbf{\Xi}_1\boldsymbol{\Psi }_{1}((\mathbf{H}_1+{\mathbf{G}^H}\boldsymbol{\Psi }_{2}\mathbf{H}_2){{\mathbf{x}}}+{\mathbf{G}^H}\boldsymbol{\Psi }_{2}{{\mathbf{n}_{2}}}+\mathbf{n}_{1}) ,\label{y_1_fd}\\
	{{\mathbf{y}_{2}}} &=\mathbf{\Xi}_2\boldsymbol{\Psi }_{2}((\mathbf{H}_2+{\mathbf{G}}\boldsymbol{\Psi }_{1}\mathbf{H}_1){{\mathbf{x}}}+{\mathbf{G}}\boldsymbol{\Psi }_{1}{{\mathbf{n}_{1}}}+\mathbf{n}_{2}).\label{y_2_fd}
\end{align}
\end{subequations}

The transmit power constraints of active RIS 1 and active RIS 2 can respectively be expressed as 
\begin{subequations}\label{transmit_power}
\begin{align} 
	&P_{1}(\mathbf{W},\boldsymbol{\Psi }_{1},\boldsymbol{\Psi }_{2})=\mathbb{E} \{ \left \| {{\mathbf{y}_{1}}}  \right \|^2_2   \}\nonumber\\&=\| \mathbf{\Xi}_1\boldsymbol{\Psi }_{1}(\mathbf{H}_1+{\mathbf{G}^H}\boldsymbol{\Psi }_{2}\mathbf{H}_2){{\mathbf{W}}}\|_{F}^{2}+\sigma _{1}^{2}\left\| \mathbf{\Xi}_1\boldsymbol{\Psi }_{1} \right\|_{F}^{2} \nonumber\\&+\sigma _{2}^{2}\| \mathbf{\Xi}_1\boldsymbol{\Psi }_{1}{\mathbf{G}^H}\boldsymbol{\Psi }_{2} \|_{F}^{2}\leq P_{1}^{\textrm{max}} ,\\
	&P_{2}(\mathbf{W},\boldsymbol{\Psi }_{1},\boldsymbol{\Psi }_{2})=\mathbb{E} \{ \left \| {{\mathbf{y}_{2}}}  \right \|^2_2   \}\nonumber\\&=\left\| \mathbf{\Xi}_2\boldsymbol{\Psi }_{2}(\mathbf{H}_2+{\mathbf{G}}\boldsymbol{\Psi }_{1}\mathbf{H}_1){{\mathbf{W}}}\right\|_{F}^{2}+\sigma _{2}^{2}\left\| \mathbf{\Xi}_2\boldsymbol{\Psi }_{2} \right\|_{F}^{2}\nonumber\\&+\sigma _{1}^{2}\left\| \mathbf{\Xi}_2\boldsymbol{\Psi }_{2}{\mathbf{G}}\boldsymbol{\Psi }_{1} \right\|_{F}^{2}\leq P_{2}^{\textrm{max}},
\end{align}
\end{subequations}
where $P_{{l}}^{\textrm{max}}$ denotes the maximum transmit power  of  the $l$-th active RIS, $ \forall \ l\in \mathcal{L}$. 
\subsection{Time Required for System Stabilization}
\begin{figure}
	\centering
	\includegraphics[width=3.2in]{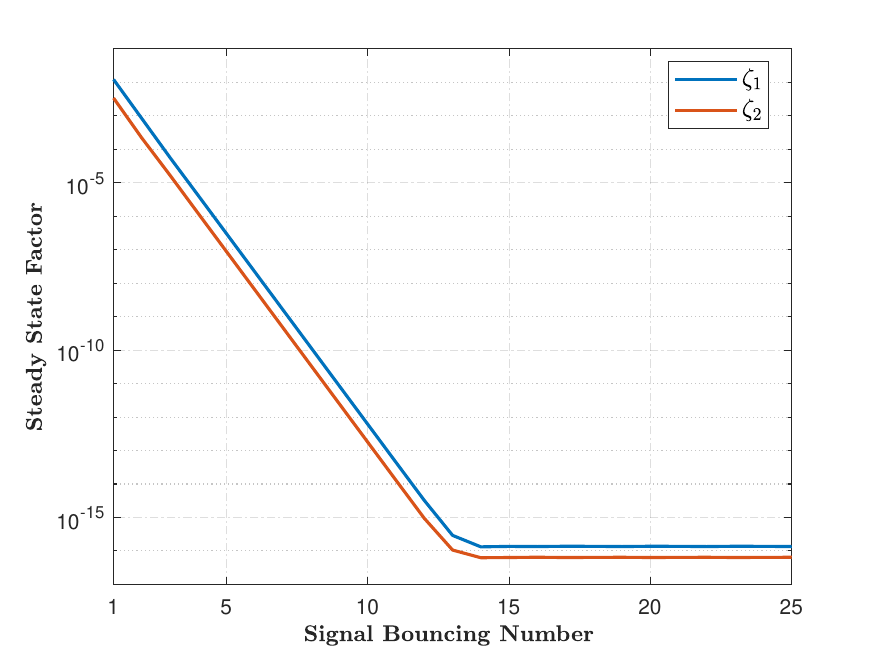}
	\caption{Steady state factor versus the number of signal bouncing. }
	\label{Plot_Steady_State}
\end{figure}

In this subsection, we discuss the time required for the signal to reach a stable state. We denote the steady state factor $\zeta_l =\left \|\tilde{\mathbf{y}}_{l} -{\mathbf{y}_{l}}  \right \|_2$, $ \forall \ l\in \mathcal{L}$, as a metric of the degree of stability of the RIS.\footnote{We illustrate the relationship between the average steady state factor and the number of signal bounces based on 100 independent channel realizations, where a random beamforming within the feasible domain is utilized, as shown in Fig. \ref{Plot_Steady_State}.} Their expressions are respectively provided in equations (\ref{instantaneous_y}) and (\ref{stabilized_y}).
It can be observed that the signal reaches a steady state after less than 20 bounces between the two RISs.
In general, the inter-excitation effect is more pronounced when two active RISs are located less than 50 meters apart, during which the time required to establish the steady state can be estimated to be approximately 5 µs.
Since time required for steady state establishment is significantly smaller than the duration of a time slot in the communication system\cite{6581606}, i.e., 577µs, time consumption for steady state establishment is negligible.
Consequently, all subsequent studies will be predicated on the assumption that these two active RISs have achieved stabilization.

\subsection{Received Signal Model}
The received signal at the $k$-th user is given by
\begin{align} \label{receivesignal}
	y_k=&\mathbf{h}_{2,k}^{H}{\mathbf{y}_{2}}+\mathbf{h}_{1,k}^{H}{\mathbf{y}_{1}}+n_k \\
	=&\bar{\mathbf{h}}_{k}^{H} {{\mathbf{w}_k}s_k}+\sum_{i=1,i\neq k}^{K}\bar{\mathbf{h}}_{k}^{H}{\mathbf{w}_i}s_i+\bar{\mathbf{g}}_{1,k}^{H}\mathbf{n}_{1}+\bar{\mathbf{g}}_{2,k}^{H}\mathbf{n}_{2}+n_k,\nonumber
\end{align}
where $\bar{\mathbf{h}}_{k}^{H}$, $\bar{\mathbf{g}}_{1,k}^{H}$ and $\bar{\mathbf{g}}_{2,k}^{H}$ are the equivalent channel respectively given by
\begin{subequations}\label{bar_h}
\begin{align}
	\bar{\mathbf{h}}_{k}^{H}&\triangleq\mathbf{h}_{1,k}^{H}\mathbf{\Xi}_1\boldsymbol{\Psi }_{1}(\mathbf{H}_1+{\mathbf{G}^H}\boldsymbol{\Psi }_{2}\mathbf{H}_2)\nonumber\\
	&+\mathbf{h}_{2,k}^{H}\mathbf{\Xi}_2\boldsymbol{\Psi }_{2}(\mathbf{H}_2+{\mathbf{G}}\boldsymbol{\Psi }_{1}\mathbf{H}_1),\\
	\bar{\mathbf{g}}_{1,k}^{H}&\triangleq\mathbf{h}_{1,k}^{H}\mathbf{\Xi}_1\boldsymbol{\Psi }_{1}+\mathbf{h}_{2,k}^{H}\mathbf{\Xi}_2\boldsymbol{\Psi }_{2}{\mathbf{G}}\boldsymbol{\Psi }_{1},\\
	\bar{\mathbf{g}}_{2,k}^{H}&\triangleq\mathbf{h}_{1,k}^{H}\mathbf{\Xi}_1\boldsymbol{\Psi }_{1}{\mathbf{G}^H}\boldsymbol{\Psi }_{2}+\mathbf{h}_{2,k}^{H}\mathbf{\Xi}_2\boldsymbol{\Psi }_{2},
\end{align}
\end{subequations}
and $n_k\sim \mathcal{CN}\left ( 0,\sigma_k^{2} \right )$ is the additive white Gaussian noise (AWGN) at the $k$-th user with zero mean and the noise power of $\sigma_k^{2}$.

Then, signal-to-interference-plus-noise ratio (SINR) 
at the $k$-th user is written as
\begin{align} 
	&\textrm{SINR}_{k}\nonumber\\
	&=\frac{|  \bar{\mathbf{h}}_{k}^{H}{\mathbf{w}_k}|^{2}}{\sum_{i=1,i\neq k}^{K}|\bar{\mathbf{h}}_{k}^{H}{\mathbf{w}_i}|^{2}+\sigma _{1}^{2}\| \bar{\mathbf{g}}_{1,k}^{H} \|_{2}^{2}+\sigma _{2}^{2}\| \bar{\mathbf{g}}_{2,k}^{H}\|^{2} +\sigma_k^{2}},
\end{align}
and the achievable rate (bit/s/Hz) of the $k$-th user is given by 
\begin{equation} 
	R_{k}=\textrm{log}_{2}\left ( 1+\textrm{SINR}_{k} \right ).
\end{equation}
\subsection{Problem Formulation}

In this paper, we maximize the WSR of users  by jointly optimizing the beamformer at the BS and the reflecting coefficients at active RISs, subject to the transmit power constraints of the BS and both active RISs, and the amplification gain constraints for each reflecting coefficients. Thus, the WSR maximization problem can be formulated as
\begin{subequations}\label{PF}
	\begin{align}
		\mathop {\max }\limits_{{\mathbf{W}, \boldsymbol{\Psi }_{1}, \boldsymbol{\Psi }_{2} } } \quad
		& \sum_{k=1}^{K} \alpha_k R_{k} \label{WSR}
		\\
		\ \textrm{s.t.}\ \quad
		& \left \| {\mathbf{W}}  \right \|_F ^2 \leq P_{\textrm{BS}}^{\textrm{max}},\label{P_BS}\\
		&P_{1}(\mathbf{W},\boldsymbol{\Psi }_{1},\boldsymbol{\Psi }_{2})\leq P_{1}^{\textrm{max}},\label{P_RIS1}\\
		&P_{2}(\mathbf{W},\boldsymbol{\Psi }_{1},\boldsymbol{\Psi }_{2})\leq P_{2}^{\textrm{max}} \label{P_RIS2}, \\
		& 1\le \left | \psi_{l,m_l} \right |^2 \le a^2_{\textrm{max}},  \ l\in \mathcal{L}, \ m_l\in \mathcal{M}_l,\label{max_a}
	\end{align}
\end{subequations}
where $\alpha_k$ is the weighting factor of the $k$-th user. 

Problem (\ref{PF}) is hard to solve directly since the objective function and constraints are non-convex and the variables are highly coupled.
Specifically, the objective function and power constraints of the active RISs include the inter-excitation matrices $\mathbf{\Xi}_1$ and $\mathbf{\Xi}_2$, and the constraint (\ref{max_a}) contribute to the non-convexity of Problem (\ref{PF}), which exacerbates the difficulty of solving Problem (\ref{PF}).

\section{Joint Beamforming Design for Double-Active-RIS-Aided Multi-User System }
In this section, we propose an algorithm to address the aforementioned challenges.
Firstly, we transform the objective function (\ref{WSR}) into a more tractable form by adopting the closed-form FP method\cite{8314727}.
Then, we introduce auxiliary variables and equality constraints to facilitate the handling of the objective function and power constraints associated with the $\mathbf{\Xi}_1$ and $\mathbf{\Xi}_2$.
Subsequently, we incorporate the equality constraints into the objective function as penalty terms by adopting the double-loop PDD algorithm \cite{9120361,9119203}.
Subsequently,  we incorporate the equality constraints as penalty terms in the objective function by adopting the double-loop PDD algorithm \cite{9120361}.
Thus, the problem can be formulated as an augmented Lagrangian (AL) problem.
Specifically, in the outer loop, we update the dual variables and the penalty parameter.
In the inner loop, we adopt the AO algorithm to decouple the converted AL problem into subproblems, which can be solved by the low-complexity  method such as the ellipsoid method and bisection based search method.

\subsection{Problem Reformulation}
By employing the FP method \cite{8314727,8310563} and introducing auxiliary variables $ \boldsymbol{{\gamma}}=\left [ {\gamma}_1 , \cdots , {\gamma}_K \right ]\in\mathbb{C}^{K}$, the  objective function (\ref{WSR}) can be equivalently reformulated as
\begin{align}
	f_{1} ( \boldsymbol{{\gamma}}, \mathbf{W}&, \boldsymbol{\Psi }_{1}, \boldsymbol{\Psi }_{2} ) =\sum_{k=1}^{K} \alpha_k\textrm{log}_{2}\left ( 1+{\gamma}_{k} \right )-\sum_{k=1}^{K} \alpha_k\frac{{\gamma}_{k}}{\textrm{ln}2}\nonumber\\
	& +\frac{1}{\textrm{ln}2} \underbrace {\sum_{k=1}^{K} \alpha_k\frac{(1+{\gamma}_{k})\textrm{SINR}_k}{(1+\textrm{SINR}_k)}}_{f_2( \boldsymbol{{\gamma}}, \mathbf{W}, \boldsymbol{\Psi }_{1}, \boldsymbol{\Psi }_{2} )}. 
\end{align}

Subsequently, the $f_2( \boldsymbol{{\gamma}}, \mathbf{W}, \boldsymbol{\Psi }_{1}, \boldsymbol{\Psi }_{2} )$ in the sum-of-ratio form can be transformed into  a quadratic function by introducing the the auxiliary variables $ \boldsymbol{\xi}=\left [ \xi_{1} , \cdots , \xi_{K} \right ]^T\in\mathbb{C}^{K}$ and adopting the quadratic transform in \cite{8314727}  as follows
\begin{align} 
	&f_3\left ( {\boldsymbol{\xi}, \boldsymbol{{\gamma}}, \mathbf{W}, \boldsymbol{\Psi }_{1}, \boldsymbol{\Psi }_{2} } \right )\nonumber\\&=2\sum_{k=1}^{K}\sqrt{\alpha_k\left ( 1+{\gamma}_k  \right )}\textrm{Re}\left\{\xi_{k}^*\bar{\mathbf{h}}_{k}^{H}{\mathbf{w}_k} \right\}-\sum_{k=1}^{K}\left| \xi_{k} \right|^2\sigma_k^{2} \nonumber
	\\&-\sum_{k=1}^{K}\left| \xi_{k} \right|^2 (\sigma _{1}^{2}\| \bar{\mathbf{g}}_{1,k} \|_{2}^{2}+\sigma _{2}^{2}\| \bar{\mathbf{g}}_{2,k} \|_{2}^{2} +\sum_{i=1}^{K} |\bar{\mathbf{h}}_{k}^{H}{\mathbf{w}_i}   |^{2}  ).
\end{align}
It can be observed that $f_3\left ( {\boldsymbol{\xi}, \boldsymbol{{\gamma}}, \mathbf{W}, \boldsymbol{\Psi }_{1}, \boldsymbol{\Psi }_{2} } \right )$ is non-convex. This is because $f_3\left ( {\boldsymbol{\xi}, \boldsymbol{{\gamma}}, \mathbf{W}, \boldsymbol{\Psi }_{1}, \boldsymbol{\Psi }_{2} } \right )$ is a function of $\boldsymbol{\Xi }_{1}$ and $\boldsymbol{\Xi }_{2}$, which are inverse matrices of functions with respect to (w.r.t.)  $\boldsymbol{\Psi }_{1}$ and $\boldsymbol{\Psi }_{2}$.
Moreover, the constraint $ \left | \psi_{l,m_l} \right |^2\ge1$ in (\ref{max_a}) is non-convex.
To address the above difficulties, we introduce the auxiliary  matrices $\mathbf{X}=\boldsymbol{\Xi }_{1}$, $\mathbf{Y}=\boldsymbol{\Xi }_{2}$ and  $\boldsymbol{\Phi }_{l} =\boldsymbol{\Psi}_l= \textrm{Diag}( \phi_{l,1},\cdots, \phi_{l,M_l})$, where $1\le \left | \phi_{l,m_l} \right |^2 \le a^2_{\textrm{max}}, \ \forall m_l\in \mathcal{M}_l \ , \forall l\in \mathcal{L}$. Then, we incorporate them as equation constraints into Problem (\ref{PF}) and transform the equation constraints with the inter-excitation matrices in a more tractable form as follows:
\begin{align}
	\mathbf{I}_{M_1}=\boldsymbol{\Xi }_{1}^{-1}\mathbf{X}=(\mathbf{I}_{M_1}-\boldsymbol{\Psi }_{1}{\mathbf{G}^H}\boldsymbol{\Psi }_{2}{\mathbf{G}})\mathbf{X}, \label{Xi_1_penalty} \\ \mathbf{I}_{M_2}=\boldsymbol{\Xi }_{2}^{-1}\mathbf{Y}=(\mathbf{I}_{M_2}-\boldsymbol{\Psi }_{2}{\mathbf{G}}\boldsymbol{\Psi }_{1}{\mathbf{G}^H})\mathbf{Y}. \label{Xi_2_penalty}
\end{align}

By defining $\mathbb{X}\triangleq\{\boldsymbol{{\gamma}}, \boldsymbol{\xi}, \mathbf{W}, \boldsymbol{\Psi }_{1}, \boldsymbol{\Psi }_{2},  \boldsymbol {\Phi }_{1}, \boldsymbol {\Phi }_{2}, \mathbf{X} , \mathbf{Y}\}$ as the set of the  optimizing variables, Problem (\ref{PF}) can be reformulated as
\begin{subequations}\label{PF_E}
\begin{align}
	\mathop {\max }\limits_{{\mathbb{X}  } } \quad
	& f_r ( \mathbb{X}   )=\sum_{k=1}^{K} \alpha_k\textrm{log}_{2}\left ( 1+{\gamma}_{k} \right ) -\frac{1}{\textrm{ln}2}\sum_{k=1}^{K} \alpha_k{\gamma}_{k} \nonumber\\
	& +\frac{2}{\textrm{ln}2}\sum_{k=1}^{K}\sqrt{\alpha_k\left ( 1+{\gamma}_k  \right )}\textrm{Re}\left\{\xi_{k}^*\tilde{\mathbf{h}}_{k}^{H}{\mathbf{w}_k} \right\}\nonumber\\
	& -\frac{1}{\textrm{ln}2}\sum_{k=1}^{K}\left| \xi_{k} \right|^2 (\sum_{i=1}^{K} |\tilde{\mathbf{h}}_{k}^{H}{\mathbf{w}_i} |^{2}+\sigma _{1}^{2}\| \tilde{\mathbf{g}}_{1,k} \|_{2}^{2}  ) \nonumber
	\\
	& -\frac{1}{\textrm{ln}2}\sum_{k=1}^{K}\left| \xi_{k} \right|^2 (\sigma _{2}^{2}\| \tilde{\mathbf{g}}_{2,k} \|_{2}^{2} +\sigma_k^{2}  ) \\
	\ \textrm{s.t.}\ \quad
	& \left \| {\mathbf{W}}  \right \|_F ^2 \leq P_{\textrm{BS}}^{\textrm{max}},\label{P_BS_E}\\
	&P_{1}(\mathbf{W},\boldsymbol{\Psi }_{1},\boldsymbol{\Psi }_{2},\mathbf{X},\mathbf{Y})\leq P_{1}^{\textrm{max}},\label{P_RIS1_E}\\
	&P_{2}(\mathbf{W},\boldsymbol{\Psi }_{1},\boldsymbol{\Psi }_{2},\mathbf{X},\mathbf{Y})\leq P_{2}^{\textrm{max}} \label{P_RIS2_E}, \\
	& (\ref{Xi_1_penalty}),\ (\ref{Xi_2_penalty}),\ \boldsymbol{\Phi }_{1}=\boldsymbol{\Psi }_{1}, \ \boldsymbol{\Phi }_{2}=\boldsymbol{\Psi }_{2}, \label{penalty_constraint}\\
	& 1\le \left | \phi_{l,m_l} \right |^2 \le a^2_{\textrm{max}},  \ l\in \mathcal{L}, \ m_l\in \mathcal{M}_l,\label{max_a_E}
\end{align}
\end{subequations}
where  $\tilde{\mathbf{h}}_{k}^{H}$, $\tilde{\mathbf{g}}_{1,k}^{H}$ $\tilde{\mathbf{g}}_{2,k}^{H}$ are respectively defined as
\begin{subequations}\label{tilde_h}
	\begin{align}
		\tilde{\mathbf{h}}_{k}^{H}&\triangleq\mathbf{h}_{1,k}^{H}\mathbf{X}\boldsymbol{\Psi }_{1}(\mathbf{H}_1+{\mathbf{G}^H}\boldsymbol{\Psi }_{2}\mathbf{H}_2)\nonumber\\
		&+\mathbf{h}_{2,k}^{H}\mathbf{Y}\boldsymbol{\Psi }_{2}(\mathbf{H}_2+{\mathbf{G}}\boldsymbol{\Psi }_{1}\mathbf{H}_1),\\
		\tilde{\mathbf{g}}_{1,k}^{H}&\triangleq\mathbf{h}_{1,k}^{H}\mathbf{X}\boldsymbol{\Psi }_{1}+\mathbf{h}_{2,k}^{H}\mathbf{Y}\boldsymbol{\Psi }_{2}{\mathbf{G}}\boldsymbol{\Psi }_{1},\\
		\tilde{\mathbf{g}}_{2,k}^{H}&\triangleq\mathbf{h}_{1,k}^{H}\mathbf{X}\boldsymbol{\Psi }_{1}{\mathbf{G}^H}\boldsymbol{\Psi }_{2}+\mathbf{h}_{2,k}^{H}\mathbf{Y}\boldsymbol{\Psi }_{2}.
	\end{align}
\end{subequations}
In (\ref{PF_E}), $P_{1}(\mathbf{W},\boldsymbol{\Psi }_{1},\boldsymbol{\Psi }_{2},\mathbf{X},\mathbf{Y})$ and $P_{2}(\mathbf{W},\boldsymbol{\Psi }_{1},\boldsymbol{\Psi }_{2},\mathbf{X},\mathbf{Y})$ represent the expressions for the transmit power of RIS 1 and RIS 2 after substituting $\mathbf{X}=\boldsymbol{\Xi }_{1}$ and $\mathbf{Y}=\boldsymbol{\Xi }_{2}$ into (\ref{transmit_power}).

For the treatment of the equality constraints  (\ref{penalty_constraint}) originating from auxiliary variables, we incorporate them into the objective function as penalty terms using the PDD algorithm \cite{9120361}. Thus, the integrated augmented Lagrangian (AL) problem can be represented as follows
\begin{subequations}\label{PF_PDD}
	\begin{align}
	\mathop {\max }\limits_{{\mathbb{X}  } } \quad & f_r (\mathbb{X})-h( \boldsymbol{\Psi }_{1}, \boldsymbol{\Psi }_{2},\boldsymbol {\Phi }_{1}, \boldsymbol {\Phi }_{2},\mathbf{X},\mathbf{Y}) \label{WSR_PDD}
		\\
		\ \textrm{s.t.}\ \quad
		& (\textrm{\ref{P_BS_E}}),(\textrm{\ref{P_RIS1_E}}),(\textrm{\ref{P_RIS2_E}}),(\textrm{\ref{max_a_E}}), \nonumber
	\end{align}
\end{subequations}
where the function $h( \boldsymbol{\Psi }_{1}, \boldsymbol{\Psi }_{2},\boldsymbol {\Phi }_{1}, \boldsymbol {\Phi }_{2},\mathbf{X},\mathbf{Y})$ is defined as
\begin{align}\label{H_ALL}
	&h( \boldsymbol{\Psi }_{1}, \boldsymbol{\Psi }_{2},\boldsymbol {\Phi }_{1}, \boldsymbol {\Phi }_{2},\mathbf{X},\mathbf{Y})\nonumber\\
	&=\frac{1}{2\rho}  \left \| \boldsymbol{\Xi }_{1}^{-1}\mathbf{X}-\mathbf{I}_{M_1}+\rho\boldsymbol{\Gamma}_1 \right \| _F^2+\frac{1}{2\rho}  \left \| \boldsymbol{\Xi }_{2}^{-1}\mathbf{Y}-\mathbf{I}_{M_2}+\rho\boldsymbol{\Gamma}_2 \right \| _F^2\nonumber\\
	&+\frac{1}{2\rho}  \left \|\boldsymbol{\psi }_{1}-\boldsymbol{\phi }_{1}+ \rho\boldsymbol{\eta}_1 \right \| _F^2+\frac{1}{2\rho}  \left \|\boldsymbol{\psi }_{2}-\boldsymbol{\phi }_{2}+ \rho\boldsymbol{\eta}_2 \right \| _F^2,
\end{align}
where $\{\boldsymbol{\Gamma}_1,\boldsymbol{\Gamma}_2,\boldsymbol{\eta}_1,\boldsymbol{\eta}_2\}$ are the dual variables, $\boldsymbol{\psi }_{l}=\textrm{diag}(\boldsymbol{\Psi }_{l}^{H})$, $\boldsymbol{\phi }_{l}=\textrm{diag}(\boldsymbol{\Phi }_{l}^{H})$, and $\rho$ is the penalty parameter. 

The adopted PDD framework \cite[TABLE II]{9120361} has double loops.
In the $t$-th outer loop iteration, the associated dual variables and the penalty parameter $\{\boldsymbol{\Gamma}_1,\boldsymbol{\Gamma}_2,\boldsymbol{\eta}_1,\boldsymbol{\eta}_2, \rho\}$ are updated according to the following expressions
\begin{subequations}\label{dual_v_rho}
	\begin{align}
		\boldsymbol{\Gamma}_1^{(t+1)}&=\boldsymbol{\Gamma}_1^{(t)}+\frac{1}{\rho^{(t)}}((\boldsymbol{\Xi }_{1}^{(t)})^{-1}\mathbf{X}^{(t)}-\mathbf{I}_{M_1}), \\
		\boldsymbol{\Gamma}_2^{(t+1)}&=\boldsymbol{\Gamma}_2^{(t)}+\frac{1}{\rho^{(t)}}((\boldsymbol{\Xi }_{2}^{(t)})^{-1}\mathbf{Y}^{(t)}-\mathbf{I}_{M_2}), \\
		\boldsymbol{\eta}_1^{(t+1)}&=\boldsymbol{\eta}_1^{(t)}+\frac{1}{\rho^{(t)}}(\boldsymbol{\psi }_{1}-\boldsymbol{\phi }_{1}), \\
		\boldsymbol{\eta}_2^{(t+1)}&=\boldsymbol{\eta}_2^{(t)}+\frac{1}{\rho^{(t)}}(\boldsymbol{\psi }_{2}-\boldsymbol{\phi }_{2}), \\
		\rho^{(t+1)}&=c\rho^{(t)},
	\end{align}
\end{subequations}
where $c$ is a constant from $0<c<1$. 
In order to measure the degree of violation of the equation constraints in each outer loop iteration, we denote $P(\mathbb{X})$ by the constraint violation indicator, which can be written as follows
\begin{align}
	P(\mathbb{X})&=\textrm{max}\{   \| \boldsymbol{\Xi }_{1}^{-1}\mathbf{X}-\mathbf{I}_{M_1} \| _{\infty},   \| \boldsymbol{\Xi }_{2}^{-1}\mathbf{Y}-\mathbf{I}_{M_2} \| _{\infty}, \nonumber\\
	& \|\boldsymbol{\psi }_{1}-\boldsymbol{\phi }_{1}  \|_{\infty}, \|\boldsymbol{\psi }_{2}-\boldsymbol{\phi }_{2} \|_{\infty}\}.
\end{align}

In the inner loop, $\{\boldsymbol{\Gamma}_1,\boldsymbol{\Gamma}_2,\boldsymbol{\eta}_1,\boldsymbol{\eta}_2, \rho\}$  is treated  as constants, and we only focus on the optimization of $\mathbb{X}$.
It is observed that Problem (\ref{PF_PDD}) is marginal convex w.r.t. $\boldsymbol{{\gamma}}, \boldsymbol{\xi}, \mathbf{W}, \boldsymbol{\Psi }_{1}, \boldsymbol{\Psi }_{2}, \mathbf{X}$ and $\mathbf{Y}$ 
individually. 
Furthermore, the closed-form optimal solution can be obtained w.r.t. $\boldsymbol {\Phi }_{1}$ and $\boldsymbol {\Phi }_{2}$ in Problem (\ref{PF_PDD}).
Therefore, we utilize the AO algorithm to solve  the AL Problem (\ref{PF_PDD})  in the following subsections.
\subsection{Optimize Beamforming Matrix $\mathbf{W}$}
In this subsection, we optimize the beamforming matrix $\mathbf{W}$ with other variables fixed. 
By removing the terms not related to $\mathbf{W}$, the subproblem of optimizing the beamforming matrix $\mathbf{W}$ can be reformulated as
\begin{subequations}\label{C_W}
	\begin{align}
		\mathop {\min }\limits_{{\mathbf{W}} } \quad
		& \textrm{Tr}\left\{\mathbf{W}^H\mathbf{A}_{\mathbf{w}}\mathbf{W}\right\}-2\textrm{Re}\left\{\textrm{Tr}\left\{\mathbf{B}_{\mathbf{w}}^{H}\mathbf{W} \right\}\right\}
		\\
		\ \textrm{s.t.}\quad
		&\textrm{Tr}\left\{\mathbf{W}^H\mathbf{W}\right\}\leq P_{\textrm{BS}}^{\textrm{max}},\label{P_w_1}\\
		&\textrm{Tr}\left\{\mathbf{W}^H\mathbf{C}_{\mathbf{w}}\mathbf{W}\right\}\leq \hat{P}_{1}^{\textrm{max}},\label{P_w_2}\\
		&\textrm{Tr}\left\{\mathbf{W}^H\mathbf{D}_{\mathbf{w}}\mathbf{W}\right\}\leq \hat{P}_{2}^{\textrm{max}}, \label{P_w_3}
	\end{align}
\end{subequations}
where the parameters $\{\mathbf{A}_{\mathbf{w}},\mathbf{B}_{\mathbf{w}},\mathbf{C}_{\mathbf{w}},\mathbf{D}_{\mathbf{w}},\hat{P}_{1}^{\textrm{max}},\hat{P}_{2}^{\textrm{max}}\}$ are respectively given by
\begin{subequations}\label{O_W_Parameter}
	\begin{align} 
		\mathbf{A}_{\mathbf{w}}&=\tilde{\mathbf{H}}\mathbf{\Lambda}\tilde{\mathbf{H}}^{H},\ {\mathbf{B}_{\mathbf{w}}}=\left [\mathbf{b}_1,\cdots,\mathbf{b}_K  \right ],\\
		{\mathbf{C}_{\mathbf{w}}}&=\mathbf{U}^H\mathbf{U}, \ {\mathbf{D}_{\mathbf{w}}}=\mathbf{V}^H\mathbf{V},\\
		\hat{P}_{1}^{\textrm{max}}&= P_{1}^{\textrm{max}}-\sigma_{1}^{2}\left\|\boldsymbol{\Xi }_{1}\boldsymbol{\Psi }_{1} \right\|_{F}^{2}-\sigma_{2}^{2}\left\|\boldsymbol{\Xi }_{1}\boldsymbol{\Psi }_{1}\mathbf{G}^T\boldsymbol{\Psi }_{2} \right\|_{F}^{2},\\
		\hat{P}_{2}^{\textrm{max}}&= P_{2}^{\textrm{max}}-\sigma_{2}^{2}\left\|\boldsymbol{\Xi }_{2}\boldsymbol{\Psi }_{2} \right\|_{F}^{2}-\sigma_{1}^{2}\left\|\boldsymbol{\Xi }_{2}\boldsymbol{\Psi }_{2}\mathbf{G}\boldsymbol{\Psi }_{1} \right\|_{F}^{2}.
	\end{align}
\end{subequations}
In (\ref{O_W_Parameter}), $\tilde{\mathbf{H}}= [\tilde{\mathbf{h}}_1,\cdots,\tilde{\mathbf{h}}_K   ]$, $\mathbf{\Lambda}=\textrm{Diag}(|\xi_{1}|^2,\cdots,|\xi_{K}|^2)$, $\mathbf{b}_k =\sqrt{\alpha_k\left ( 1+{\gamma}_k  \right )}\xi_{k}\tilde{\mathbf{h}}_{k}$, $\mathbf{U}=\mathbf{\Xi}_1\mathbf{\Psi}_1(\mathbf{H}_1+\mathbf{G}^H\mathbf{\Psi}_2\mathbf{H}_2)$, $\mathbf{V}=\mathbf{\Xi}_2\mathbf{\Psi}_2(\mathbf{H}_2+\mathbf{G}\mathbf{\Psi}_1\mathbf{H}_1)$.

It is readily derived that Problem (\ref{C_W}) is a standard convex second order cone programming (SOCP) problem, which can be solved efficiently via standard optimization packages, e.g. CVX\cite{grant2014cvx}. 
However, the computational complexity of employing CVX to solve an SOCP problem is excessive. 

To reduce the computational complexity, we employ the ellipsoid method \cite{boydEE392} to solve the dual form of Problem (\ref{C_W}).

%

The Lagrangian function of Problem (\ref{C_W}) can be represented as 
%
%
\begin{align}
	&\mathcal{L}(\mathbf{W},\boldsymbol{\lambda})=\textrm{Tr}\left\{\mathbf{W}^H\mathbf{A}_{\mathbf{w}}\mathbf{W}\right\}-2\textrm{Re}\{\mathbf{B}_{\mathbf{w}}^{H}\mathbf{W}\}\nonumber\\
	&+\lambda_1 ( \textrm{Tr}\left\{\mathbf{W}^H\mathbf{W}\right\}- P_{\textrm{BS}}^{\textrm{max}} )+\lambda_2(\textrm{Tr}\left\{\mathbf{W}^H\mathbf{C}_{\mathbf{w}}\mathbf{W}\right\}-\hat{P}_{1}^{\textrm{max}} )\nonumber\\
	&+\lambda_3(\textrm{Tr}\left\{\mathbf{W}^H\mathbf{D}_{\mathbf{w}}\mathbf{W}\right\}-\hat{P}_{2}^{\textrm{max}}),
\end{align}
where $\boldsymbol{\lambda}=\left [\lambda_1,\lambda_2, \lambda_3  \right ]^T  \succeq \mathbf{0} $ is the Lagrange multiplier vector. Then, the dual problem of (\ref{C_W}) can be expressed as
\begin{subequations}\label{L_W}
	\begin{align}
		\mathop {\max }\limits_{{\boldsymbol{\lambda}} }& \ 
		 g({\boldsymbol{\lambda}})= \mathop {\min }\limits_{{\mathbf{W}}\in \mathcal{D}_{\mathbf{w}} } \ \mathcal{L}(\mathbf{W},\boldsymbol{\lambda}) \label{L_W_O}
		\\
		\ \textrm{s.t.}&\
		\boldsymbol{\lambda}\succeq \mathbf{0},
	\end{align}
\end{subequations}
where the feasibility domain of $\mathbf{W}$ in Problem (\ref{C_W}) is denoted by
$\mathcal{D}_{\mathbf{w}}=\{ \mathbf{W} |  (\textrm{\ref{P_w_1}}) \cap (\textrm{\ref{P_w_2}}) \cap (\textrm{\ref{P_w_3}})   \}$.
By setting $\partial \mathcal{L}(\mathbf{W},\boldsymbol{\lambda})/\partial\mathbf{W}=0$, the optimal beamforming matrix $\mathbf{W}$ can be obtained as 
\begin{align}
	\mathbf{W}(\boldsymbol{\lambda})=\left ( \mathbf{A}_\mathbf{w}+\lambda_1\mathbf{I}_N+\lambda_2\mathbf{C}_\mathbf{w}+\lambda_3\mathbf{D}_\mathbf{w} \right )^{-1}\mathbf{B}_\mathbf{w}.
\end{align}

Note that the optimal beamforming matrix $\mathbf{W}^\textrm{opt}$ is determined by the Lagrange multiplier vector $\boldsymbol{\lambda}$, which should satisfy the  complementary slackness conditions for the constraints (\ref{P_w_1}), (\ref{P_w_2}), (\ref{P_w_3}). Thus, we search for the optimal Lagrange multiplier vector $\boldsymbol{\lambda}^\textrm{opt}$ by adopting the ellipsoid method. 

According to \cite[Section III]{boydEE392},
	we update $\mathbf{g}^{(\tau)}$ by using the subgradient of the objective function if ${\boldsymbol{\lambda}}^{(\tau)}$ is feasible; otherwise, we update $\mathbf{g}^{(\tau)}$ by using the subgradient of the constraint function corresponding to the violated constraint.
	Therefore, for Problem (\ref{L_W}), the selection of the subgradient $\mathbf{g}^{(\tau)}$ for $g({\boldsymbol{\lambda}})$ with feasible $\boldsymbol{\lambda}$ is given as
	\begin{align}\label{sub_g_tau}
		{{\mathbf{g}}^{(\tau)}}=\begin{bmatrix}
			P_{\textrm{BS}}-\textrm{Tr}\{{\mathbf{W}^{H}(\boldsymbol{\lambda})}{\mathbf{W}(\boldsymbol{\lambda})} \} \\
			\hat{P}_{1}^{\textrm{max}}-\textrm{Tr}\{{\mathbf{W}^{H}(\boldsymbol{\lambda})}\mathbf{C}_{\mathbf{w}}{\mathbf{W}(\boldsymbol{\lambda})} \}  \\
			\hat{P}_{2}^{\textrm{max}}-\textrm{Tr}\{{\mathbf{W}^{H}(\boldsymbol{\lambda})}\mathbf{D}_{\mathbf{w}}{\mathbf{W}(\boldsymbol{\lambda})} \} 
		\end{bmatrix},
	\end{align}
	which can be proved to be a subgradient of $g({\boldsymbol{\lambda}})$ similarly to the approach presented in \cite{1658226}. When $\mathbf{W}(\boldsymbol{\lambda})$ or $\boldsymbol{\lambda}$ is infeasible, the subgradient $\mathbf{g}$ is given by ${{\mathbf{g}}^{(\tau)}}= \textrm{min}\{\textrm{sgn} ( \boldsymbol{\lambda} ),\mathbf{0} \} \ \textrm{or} \ \textrm{min}\{\textrm{sgn} ( {{\mathbf{g}}^{(\tau)}} ),\mathbf{0} \}$. 
	Based on the above analysis, the detailed procedures of the ellipsoid method for solving Problem (\ref{C_W})  are given in Algorithm \ref{AO_W_M}.

\begin{algorithm}
	\caption{Ellipsoid Method for Solving Problem (\ref{C_W})}
	\label{AO_W_M}
	\begin{algorithmic}[1]
	 \REQUIRE Initialize the iteration number $\tau = 0$, the maximum iteration time for the beamforming vector optimization $\tau^{\textrm{max}}$, the dimension of the ellipsoid method $n=3$, the initial ellipsoid ($\boldsymbol{\Pi}^{(0)}$, $\boldsymbol{\lambda}^{(0)}$), where $\boldsymbol{\Pi}^{(0)}=R^2\mathbf{I}_{\boldsymbol{\lambda}}$.
	  \REPEAT
	  \STATE Calculate the beamforming vector $\mathbf{W}^{(j)}=\mathbf{W}(\boldsymbol{\lambda}^{(\tau)})$;
		\IF{$\boldsymbol{\lambda}^{(\tau)}\succeq\mathbf{0}$} 
		\STATE  the subgradient $\mathbf{g}^{(\tau)}$ is given by (\ref{sub_g_tau});
	\IF{$\mathbf{W}(\boldsymbol{\lambda}^{(\tau)})\notin\mathcal{D}$ }
	\STATE${{\mathbf{g}}^{(\tau)}}= \textrm{min}\{\textrm{sgn} ( {\mathbf{g}}^{(\tau)} ),\mathbf{0} \}$;
	\ENDIF
	   \ELSE
	   \STATE ${{\mathbf{g}}^{(\tau)}}= \textrm{min}\{\textrm{sgn} ( \boldsymbol{\lambda}^{(\tau)} ),\mathbf{0} \}$;
	   \ENDIF
		\STATE Calculate the normalized subgradient $\hat{\mathbf{g}}^{(\tau)}$ by $\qquad\qquad\hat{\mathbf{g}}^{(\tau)}=\frac{1}{\sqrt{\mathbf{g}^{(\tau)T}\boldsymbol{\Pi}^{(\tau)}\mathbf{g}^{(\tau)}}}\mathbf{g}^{(\tau)}$;
		\STATE Update ellipsoid center $\boldsymbol{\lambda}^{(\tau+1)}$ by 
		$\boldsymbol{\lambda}^{(\tau+1)}=\boldsymbol{\lambda}^{(\tau)}-\frac{1 }{n+1}\boldsymbol{\Pi}^{(\tau)}{\hat{\mathbf{g}}^{(\tau)}}$;
		\STATE Update ellipsoid shape  $\boldsymbol{\Pi}^{(\tau+1)}$ by 
		$\boldsymbol{\Pi}^{(\tau+1)}=\frac{n^2}{n^2-1}\left (\boldsymbol{\Pi}^{(\tau)}-\frac{2}{n+1}\boldsymbol{\Pi}^{(\tau)}{\hat{\mathbf{g}}^{(\tau)}}{\hat{\mathbf{g}}^{(\tau)T}}\boldsymbol{\Pi}^{(\tau)} \right )$;
		\STATE $\tau=\tau+1$;
		\UNTIL{ $\tau=\tau^{\textrm{max}}$ }
	\end{algorithmic}
\end{algorithm}
\subsection{Optimize Reflecting Coefficient Matrices $\boldsymbol{\Psi }_{1}$ and $\boldsymbol{\Psi }_{2}$ for Double Active RISs}
In this subsection, we alternately optimize $\boldsymbol{\Psi }_{1}$ and $\boldsymbol{\Psi }_{2}$  by fixing other variables.
Firstly, we optimize $\boldsymbol{\Psi }_{1}$ with fixed $\boldsymbol{\Psi }_{2}$.
By removing the terms that are unrelated to $\boldsymbol{\Psi }_{1}$, we can reformulate Problem (\ref{PF_PDD}) as follows 

\begin{subequations}\label{Psi_1}
	\begin{align}
		\mathop {\min }\limits_{ \boldsymbol{\Psi }_{1}} \quad
		& \frac{1}{\textrm{ln}2}\sum_{k=1}^{K}\left| \xi_{k} \right|^2(\textrm{Tr}(\boldsymbol{\Psi }_{1}\mathbf{E}_1\boldsymbol{\Psi }_{1}^H\mathbf{F}_{1,k})+\textrm{Tr}(\boldsymbol{\Psi }_{1}\boldsymbol{\Psi }_{1}^H\mathbf{F}_{2,k}))\nonumber\\& \frac{1}{2\rho}(\textrm{Tr}(\boldsymbol{\Psi }_{1}\mathbf{E}_3\boldsymbol{\Psi }_{1}^H\mathbf{F}_3)+\textrm{Tr}(\boldsymbol{\Psi }_{1}\mathbf{E}_4\boldsymbol{\Psi }_{1}^H))\nonumber\\&-2\textrm{Re}\{\textrm{Tr}(\boldsymbol{\Psi }_{1}\mathbf{P}_1)\}\\
		\ \textrm{s.t.}\quad
		&\textrm{Tr}(\boldsymbol{\Psi }_{1}\mathbf{E}_5\boldsymbol{\Psi }_{1}^H\mathbf{F}_5)\leq P_{1}^{\textrm{max}},\\
		&\textrm{Tr}(\boldsymbol{\Psi }_{1}\mathbf{E}_6\boldsymbol{\Psi }_{1}^H\mathbf{F}_6)+2\textrm{Re}\{\textrm{Tr}(\boldsymbol{\Psi }_{1}\mathbf{P}_2)\}\leq \tilde{P}_{2}^{\textrm{max}},
	\end{align}
\end{subequations}
where the parameters are respectively given by
\begin{subequations}\label{o_Psi_1_parameter}
	\begin{align}
		\mathbf{E}_{1}&=\tilde{\mathbf{H}}_1\mathbf{W}\mathbf{W}^H\tilde{\mathbf{H}}_1^H+\sigma _{2}^{2}\mathbf{G}^H\boldsymbol{\Psi }_{2}\boldsymbol{\Psi }_{2}^H\mathbf{G},\\
		\mathbf{F}_{1,k}&=\mathbf{X}^{H}\mathbf{h}_{1,k}\mathbf{h}_{1,k}^{H}\mathbf{X} , \ \mathbf{F}_{2,k}=\mathbf{R}_{3,k}^H\mathbf{R}_{3,k},\\
		\mathbf{E}_{3}&={\mathbf{G}^H}\mathbf{Y}\mathbf{Y}^H\mathbf{G}, \ \mathbf{F}_{3}=\mathbf{G}^H\boldsymbol{\Psi }_{2}^H\boldsymbol{\Psi }_{2}\mathbf{G},\\
		\mathbf{E}_{4}&={\mathbf{G}^H}\boldsymbol{\Psi }_{2}{\mathbf{G}}\mathbf{X}({\mathbf{G}^H}\boldsymbol{\Psi }_{2}{\mathbf{G}}\mathbf{X})^H+\mathbf{I}_{M_1},\\
		\mathbf{P}_{1}&=\frac{1}{\textrm{ln}2}\sum_{k=1}^{K}\sqrt{\alpha_k\left ( 1+{\gamma}_k  \right )}\xi_{k}^*\mathbf{R}_1\mathbf{w}_{k}\mathbf{h}_{1,k}^{H}\mathbf{X}\nonumber\\
		&-\frac{1}{\textrm{ln}2}\sum_{k=1}^{K}\left| \xi_{k} \right|^2(\mathbf{R}_1\mathbf{W}\mathbf{W}^H\mathbf{H}_2^H+\sigma _{2}^{2}\mathbf{G}^H\boldsymbol{\Psi }_{2})\mathbf{R}_{2,k}\nonumber\\
		&+\frac{1}{2\rho}{\mathbf{G}^H}\boldsymbol{\Psi }_{2}{\mathbf{G}}\mathbf{X}(\mathbf{X}-\mathbf{I}_{M_1}+\rho\boldsymbol{\Gamma}_1)^H\nonumber\\
		&+\frac{1}{2\rho}{\mathbf{G}^H}\mathbf{Y}(\mathbf{Y}-\mathbf{I}_{M_2}+\rho\boldsymbol{\Gamma}_2)^H\boldsymbol{\Psi }_{2}\mathbf{G}\nonumber\\
		&+\frac{1}{2\rho}(\boldsymbol{\Phi }_{1}-\rho\textrm{Diag}(\boldsymbol{\eta}_1^H)),\nonumber\\
		\mathbf{E}_5&=(\mathbf{H}_1+{\mathbf{G}^H}\boldsymbol{\Psi }_{2}\mathbf{H}_2){{\mathbf{W}}}{{\mathbf{W}}}^H(\mathbf{H}_1+{\mathbf{G}^H}\boldsymbol{\Psi }_{2}\mathbf{H}_2)^H\nonumber\\
		&+\sigma _{1}^{2}\mathbf{I}_{M_1}+\sigma _{2}^{2}{\mathbf{G}^H}\boldsymbol{\Psi }_{2}\boldsymbol{\Psi }_{2}^H{\mathbf{G}},\nonumber\\
		\mathbf{E}_6&=\mathbf{H}_1{{\mathbf{W}}}{{\mathbf{W}}}^H\mathbf{H}_1^H+\sigma _{1}^{2}\mathbf{I}_{M_1},\\
		\mathbf{F}_5&=\mathbf{X}^H\mathbf{X}, \ \mathbf{F}_6=(\mathbf{Y}\boldsymbol{\Psi }_{2}{\mathbf{G}})^H\mathbf{Y}\boldsymbol{\Psi }_{2}{\mathbf{G}},\\
		\mathbf{P}_2&=\mathbf{H}_1{{\mathbf{W}}}{{\mathbf{W}}}^H\mathbf{H}_2^H\boldsymbol{\Psi }_{2}^H\mathbf{Y}^H\mathbf{Y}\boldsymbol{\Psi }_{2}{\mathbf{G}},\\
		\tilde{P}_{2}^{\textrm{max}}&={P}_{2}^{\textrm{max}}-\|\mathbf{Y}\boldsymbol{\Psi }_{2}\mathbf{H}_2{{\mathbf{W}}} \|_F^2
		-\sigma _{2}^{2}\|\mathbf{Y}\boldsymbol{\Psi }_{2}\|_F^2.
	\end{align}
\end{subequations}
In (\ref{o_Psi_1_parameter}), $\mathbf{R}_1=\mathbf{H}_1+{\mathbf{G}^H}\boldsymbol{\Psi }_{2}\mathbf{H}_2$, $\mathbf{R}_{2,k}=\boldsymbol{\Psi }_{2}^H\mathbf{Y}^H\mathbf{h}_{2,k}\mathbf{h}_{1,k}^{H}\mathbf{X}$,
$\mathbf{R}_{3,k}=\mathbf{h}_{1,k}^{H}\mathbf{X}+\mathbf{h}_{2,k}^{H}\mathbf{Y}\boldsymbol{\Psi }_{2}{\mathbf{G}}$.

By using the matrix property $\textrm{Tr}(\boldsymbol{\Psi }_{1}\mathbf{A}\boldsymbol{\Psi }_{1}^H\mathbf{B})=\boldsymbol{\psi }_{1}^H(\mathbf{A}\odot\mathbf{B}^T)\boldsymbol{\psi }_{1}$ and $\textrm{Tr}(\boldsymbol{\Psi }_{1}\mathbf{C})=\boldsymbol{\psi }_{1}^H\textrm{diag}(\mathbf{C})$ in \cite{zhang2017matrix}, Problem (\ref{Psi_1}) can be reformulated as
\begin{subequations}\label{O_Psi_1_PDD_IN}
	\begin{align}
		\mathop {\min }\limits_{{ \boldsymbol{\psi }_{1}} } \quad
		&  \boldsymbol{\psi }_{1}^H\mathbf{A}_{\boldsymbol{\psi}_{1}}\boldsymbol{\psi }_{1}-2\textrm{Re}\{\boldsymbol{\psi }_{1}^H\mathbf{b}_{\boldsymbol{\psi}_{1}}\}
		\\
		\ \textrm{s.t.}\ \quad
		&\boldsymbol{\psi}_{1}^H\mathbf{C}_{\boldsymbol{\psi}_{1}}\boldsymbol{\psi}_{1}\leq P_{1}^{\textrm{max}},\label{P_RIS1_C_Psi}\\
		&\boldsymbol{\psi}_{1}^H\mathbf{D}_{\boldsymbol{\psi}_{1}}\boldsymbol{\psi}_{1}+2\textrm{Re}\{\boldsymbol{\psi}_{1}^H\mathbf{d}_{\boldsymbol{\psi}_{1}}\}+\leq \tilde{P}_{2}^{\textrm{max}} \label{P_RIS2_C_Psi}, 
	\end{align}
\end{subequations}
where
\begin{subequations}\label{o_psi_1_parameter}
	\begin{align}
		\mathbf{A}_{\boldsymbol{\psi}_{1}}&=	\frac{1}{\textrm{ln}2}\sum_{k=1}^{K}\left| \xi_{k} \right|^2(\mathbf{E}_1\odot\mathbf{F}_{1,k}^T+\mathbf{I}_{M_1}\odot\mathbf{F}_{2,k}^T)\nonumber\\&+\frac{1}{2\rho}(\mathbf{E}_3\odot\mathbf{F}_{3,k}^T+\mathbf{E}_4\odot\mathbf{I}_{M_1}),\\
		\mathbf{b}_{\boldsymbol{\psi}_{1}}&=\textrm{diag}(\mathbf{P}_1),\ \mathbf{C}_{\boldsymbol{\psi}_{1}}=\mathbf{E}_5\odot\mathbf{F}_5^T,\\
		\mathbf{D}_{\boldsymbol{\psi}_{1}}&=\mathbf{E}_6\odot\mathbf{F}_6^T, \
		\mathbf{d}_{\boldsymbol{\psi}_{1}}= \textrm{diag}(\mathbf{P}_2).
	\end{align}
\end{subequations}
It is noted that  Problem (\ref{O_Psi_1_PDD_IN})
is a convex optimization problem, which can be effectively addressed by using the aforementioned ellipsoid method. 

Due to the symmetry of $\boldsymbol{\Psi }_{1}$ and $\boldsymbol{\Psi }_{2}$, the algorithm for optimizing $\boldsymbol{\Psi }_{1}$  can also  work for optimizing $\boldsymbol{\Psi }_{2}$. Therefore, we will not go into details about the process of solving for $\boldsymbol{\Psi }_{2}$.
\subsection{Optimize Auxiliary Variables $\boldsymbol{{\gamma}}$, $\boldsymbol{\xi}$, $\boldsymbol{\Phi}_1$, $\boldsymbol{\Phi}_2$, $\mathbf{X}$ and $\mathbf{Y}$}
In this subsection, we alternately optimize the auxiliary variables $\{\boldsymbol{{\gamma}}, \boldsymbol{\xi}, \boldsymbol{\Phi}_1, \boldsymbol{\Phi}_2, \mathbf{X}, \mathbf{Y} \}$. The procedures for solving these subproblems are described as follows.

\textbf{\textit{1) Optimizing Auxiliary Vector $\boldsymbol{\xi}$:}}
By fixing other variables and setting $\partial f_r(\mathbb{X})/\partial\xi_{k}=0$,
we can obtain the optimal $\xi_{k}^{\textrm{opt}}$ as follows
\begin{align} \label{xi}
	\xi_{k}^{\textrm{opt}}=\frac{\sqrt{\alpha_k\left ( 1+{\gamma}_k  \right )}\tilde{\mathbf{h}}_{k}^{H}{\mathbf{w}_k}}{\sum_{i=1}^{K}|\tilde{\mathbf{h}}_{k}^{H}{\mathbf{w}_i}   |^{2}+\tilde{\sigma}_{k}^{2}}, \ \forall k\in\mathcal{K},
\end{align}
where $\tilde{\sigma}_{k}^{2}=\sigma _{1}^{2}\| \tilde{\mathbf{g}}_{1,k}^{H} \|_{2}^{2}+\sigma _{2}^{2}\| \tilde{\mathbf{g}}_{2,k}^{H}\|_{2}^{2} +\sigma_k^{2}$.

\textbf{\textit{2) Optimizing Auxiliary Vector $\boldsymbol{{\gamma}}$:}}
By fixing other variables, substituting (\ref{xi}) into $f_r(\mathbb{X})$, and setting $\partial f_r(\mathbb{X})/\partial\gamma_{k}=0$, the optimal $\gamma_{k}^{\textrm{opt}}$, $\forall  k \in \mathcal{K}$ can be obtained
as
\begin{align} \label{gamma}
	{\gamma}_k^{\textrm{opt}}=\frac{|  \tilde{\mathbf{h}}_{k}^{H}{\mathbf{w}_k}|^{2}}{\sum_{i=1,i\neq k}^{K}|\tilde{\mathbf{h}}_{k}^{H}{\mathbf{w}_i}|^{2}+\tilde{\sigma}_{k}^{2}}.
\end{align}

\textbf{\textit{3) Optimizing Auxiliary Matrices $\boldsymbol{\Phi }_{l}$, $\forall l\in\mathcal{L}$:}}
In this block, we optimize $\boldsymbol{\Phi }_{l}$ with other variables fixed. By removing the terms unrelated to $\boldsymbol{\Phi }_{l}$, Problem (\ref{PF_PDD}) is simplified as follows
\begin{subequations}
	\begin{align}
		\mathop {\min }\limits_{{ \boldsymbol{\phi }_{l}} } \quad & \| \boldsymbol{\psi}_l-\boldsymbol{\phi}_l+\rho\boldsymbol{\eta}_l  \| ^2_2,\\ 
		\ \textrm{s.t.}\ \quad
		&1\le  | \phi_{l,m_l}  |^2 \le a^2_{\textrm{max}}, \ \forall m_l\in \mathcal{M}_l.
	\end{align}
\end{subequations}
By using the same method in \cite{10134546}, the optimum $\boldsymbol{\phi}_l$ can be obtained as
	\begin{align}\label{phi_1}
\boldsymbol{\phi}_l= [ |\boldsymbol{\psi}_l +\rho\boldsymbol{\eta}_l |  ]^{a_{\textrm{max}}}_{1} \textrm{exp}(j\angle (\boldsymbol{\psi}_l +\rho\boldsymbol{\eta}_l)).
	\end{align}

\textbf{\textit{4) Optimizing Auxiliary Matrices $\mathbf{X}$ and $\mathbf{Y}$:}}
In this block, we focus on optimizing $\mathbf{X}$ and $\mathbf{Y}$ by
fixing other variables.
Firstly, we optimize $\mathbf{X}$ with fixed $\mathbf{Y}$.
By removing the terms irrelevant to $\mathbf{X}$, the subproblem of optimizing the auxiliary matrix $\mathbf{X}$ can be formulated as follows
\begin{subequations}\label{O_X_PDD_IN}
	\begin{align}
		\mathop {\min }\limits_{{ \mathbf{X}} } \quad
		&  \textrm{Tr}(\mathbf{X}\mathbf{M}_1\mathbf{X}^H\mathbf{L}_1)+\textrm{Tr}(\mathbf{X}\mathbf{X}^H\mathbf{L}_2)-2\textrm{Re}\{\textrm{Tr}(\mathbf{X}\mathbf{L}_3)\}
		\\
		\ \textrm{s.t.}\ \quad
		&\textrm{Tr}(\mathbf{X}\mathbf{M}_1\mathbf{X}^H)\leq P_{1}^{\textrm{max}},\label{P_RIS1_C_X}
	\end{align}
\end{subequations}
where 
\begin{subequations}\label{o_X_parameter}
	\begin{align}
		\mathbf{M}_1&=\boldsymbol{\Psi }_{1}(\mathbf{H}_1+{\mathbf{G}^H}\boldsymbol{\Psi }_{2}\mathbf{H}_2){{\mathbf{W}}}(\boldsymbol{\Psi }_{1}(\mathbf{H}_1+{\mathbf{G}^H}\boldsymbol{\Psi }_{2}\mathbf{H}_2){{\mathbf{W}}})^H\nonumber\\
		&+\sigma _{1}^{2}\boldsymbol{\Psi }_{1}\boldsymbol{\Psi }_{1}^H+\sigma _{2}^{2}\boldsymbol{\Psi }_{1}{\mathbf{G}^H}\boldsymbol{\Psi }_{2}(\boldsymbol{\Psi }_{1}{\mathbf{G}^H}\boldsymbol{\Psi }_{2})^H, \\
		\mathbf{L}_1&=\frac{1}{\textrm{ln}2}\sum_{k=1}^{K}\left| \xi_{k} \right|^2\mathbf{h}_{1,k}\mathbf{h}_{1,k}^H, \
		\mathbf{L}_2=\frac{1}{2\rho}(\boldsymbol{\Xi }_{1}^{-1})^H\boldsymbol{\Xi }_{1}^{-1},\\
		\mathbf{L}_3&=\frac{1}{\textrm{ln}2}\sum_{k=1}^{K}\sqrt{\alpha_k\left ( 1+{\gamma}_k  \right )}\xi_{k}^*\mathbf{T}_1{\mathbf{w}_k\mathbf{h}_{1,k}^H} +\frac{1}{2\rho}\boldsymbol{\Xi }_{1}^{-1}\\&-\frac{1}{\textrm{ln}2}\sum_{k=1}^{K}\left| \xi_{k} \right|^2(\mathbf{T}_1\mathbf{W}\mathbf{W}^H\mathbf{T}_2^H\mathbf{Y}^H\mathbf{h}_{2,k}\mathbf{h}_{1,k}^H\nonumber\\
		&+\sigma _{1}^{2}\boldsymbol{\Psi}_1\mathbf{T}_3^H\mathbf{Y}^H\mathbf{h}_{2,k}\mathbf{h}_{1,k}^H+\sigma _{2}^{2}\mathbf{T}_4\boldsymbol{\Psi}_2^H\mathbf{Y}^H\mathbf{h}_{2,k}\mathbf{h}_{1,k}^H).\nonumber
	\end{align}
\end{subequations}
In (\ref{o_X_parameter}), ${\mathbf{T}}_{1}=\boldsymbol{\Psi }_{1}(\mathbf{H}_1+{\mathbf{G}^H}\boldsymbol{\Psi }_{2}\mathbf{H}_2)$, $ {\mathbf{T}}_{2}=\boldsymbol{\Psi }_{2}(\mathbf{H}_2+{\mathbf{G}}\boldsymbol{\Psi }_{1}\mathbf{H}_1)$, ${\mathbf{T}}_{3}=\boldsymbol{\Psi }_{2}{\mathbf{G}}\boldsymbol{\Psi }_{1}$, $ {\mathbf{T}}_{4}=\boldsymbol{\Psi }_{1}{\mathbf{G}^H}\boldsymbol{\Psi }_{2}$.

It can be readily verified that matrix $\mathbf{M}_1$ is positive semi-definite, which can be decomposed as $\mathbf{M}_1=\mathbf{K}\mathbf{K}^H$. 
By denoting  $\hat{\mathbf{x}}=\textrm{vec}(\mathbf{X}\mathbf{K})$ and exploiting properties  $\textrm{Tr}(\mathbf{A}\mathbf{B}\mathbf{C}\mathbf{D})=(\textrm{vec}(\mathbf{D}^T))^T(\mathbf{C}^T\otimes \mathbf{A})\textrm{vec}(\mathbf{B})$, $\textrm{Tr}(\mathbf{A}^H\mathbf{B})=(\textrm{vec}(\mathbf{A}))^H\textrm{vec}(\mathbf{B})$   in \cite[1.11]{zhang2017matrix}, Problem (\ref{O_X_PDD_IN}) can be reformulated as follows
\begin{subequations}\label{O_hat_X_PDD_IN}
	\begin{align}
		\mathop {\min }\limits_{{ \hat{\mathbf{x}}} } \quad
		& \hat{\mathbf{x}}^H\mathbf{A}_{\hat{\mathbf{x}}}\hat{\mathbf{x}}--2\textrm{Re}\{\hat{\mathbf{x}}^H\mathbf{b}_{\hat{\mathbf{x}}}\} 
		\\
		\ \textrm{s.t.}\ \quad
		&\left \| {\hat{\mathbf{x}}}  \right \|_2 ^2\leq P_{1}^{\textrm{max}},\label{P_RIS1_C_hat_X}
	\end{align}
\end{subequations}
where $\mathbf{A}_{\hat{\mathbf{x}}}=\mathbf{I}_{M_1}\otimes \mathbf{L}_1+((\mathbf{K}^H\mathbf{K})^{-1})^T\otimes \mathbf{L}_2$ and $\mathbf{b}_{\hat{\mathbf{x}}}=\textrm{vec}(\hat{\mathbf{L}}_3^H)$.
It is observed that Problem (\ref{O_hat_X_PDD_IN}) is a convex problem that can be solved by CVX tool\cite{grant2014cvx}. However, using CVX tool to tackle the SOCP problem results in high computational complexity. In order to solve Problem (\ref{O_hat_X_PDD_IN}) with low computational complexity, we adopt the Lagrangian multiplier method to obtain the quasi-closed solution. 
The Lagrangian function of Problem (\ref{O_hat_X_PDD_IN}) is given by $\mathcal{L}_{x}(\hat{\mathbf{x}},\mu )=\hat{\mathbf{x}}^H\mathbf{A}_{\hat{\mathbf{x}}}\hat{\mathbf{x}}-2\textrm{Re}\{\hat{\mathbf{x}}^H\mathbf{b}_{\hat{\mathbf{x}}}\}+\mu(\hat{\mathbf{x}}^H\hat{\mathbf{x}}-P_{1}^{\textrm{max}})$. By setting $\partial \mathcal{L}_{x} (\hat{\mathbf{x}},\mu )/\partial\hat{\mathbf{x}}^\ast =\mathbf{0}$, the optimal solution of $\hat{\mathbf{x}}$ can be obtained as follows
\begin{align}
	\hat{\mathbf{x}}(\mu)=(\mathbf{A}_{\mathbf{x}}+\mu\mathbf{I}_{M_1^2})^{-1}\mathbf{b}_{\hat{\mathbf{x}}},
\end{align}
where the optimal Lagrange multiplier $\mu$ can be readily found by the bisection search method\cite{9090356}. Since the method of optimizing $\mathbf{X}$   also  works for optimizing $\mathbf{Y}$, we will not go into details about the process of solving $\mathbf{Y}$.

\subsection{Overall Algorithm to Solve Problem (\ref{PF})}
In the aforementioned discussions, the detailed description of the PDD algorithm to solve Problem (\ref{PF}) is summarized in Algorithm \ref{PDD}. Similar to the proof in \cite{9120361}, the PDD-based Algorithm \ref{PDD} can converge to a stationary point of Problem (\ref{PF}).
\begin{algorithm}
	\caption{PDD Algorithm  for Solving Problem (\ref{PF})}
	\label{PDD}
	\begin{algorithmic}[1]
		\STATE	\textbf{Require:} Initialize the iteration number $t = 1$, the maximum number of iteration $t^{\textrm{max}}$, feasible $\mathbf{W}^{\left (1  \right )}$, $\boldsymbol{\Psi}_{1}^{\left (1  \right )}$ and $\boldsymbol{\Psi}_{2}^{\left (1  \right )}$, the tolerance of accuracy $\delta$ and  $\epsilon$. 
		\STATE Calculate the initial auxiliary matrices
		 $\mathbf{X}^{\left (1  \right )}=(\boldsymbol{\Xi }_{1}^{(1)})^{-1}$, 
		 $\mathbf{Y}^{\left (1  \right )}=(\boldsymbol{\Xi }_{2}^{(1)})^{-1}$,
		 $\boldsymbol{\Phi }_{1}^{\left (1  \right )}=\boldsymbol{\Psi }_{1}^{\left (1  \right )}$, $\boldsymbol{\Phi }_{2}^{\left (1  \right )}=\boldsymbol{\Psi }_{2}^{\left (1  \right )}$; 
		\REPEAT 
		\REPEAT
		\STATE Calculate the optimal auxiliary vector $\boldsymbol{\gamma}$ in (\ref{gamma}) with other variables fixed;
		\STATE Calculate the optimal auxiliary vector $\boldsymbol{\xi}$ according to (\ref{xi}) with other variables fixed;
		\STATE Calculate the value of  objective function of Problem (\ref{PF_PDD});
		\STATE Calculate the optimal beamforming matrix $\mathbf{W}$ by solving Problem (\ref{C_W}) using ellipsoid method in Subsection III-B with other variables fixed;
		\STATE Calculate the optimal reflecting coefficient matrix $\boldsymbol{\Psi}_{1}$ by solving Problem (\ref{Psi_1}) using  ellipsoid method with other variables fixed;
		\STATE Calculate the optimal reflecting coefficient matrix $\boldsymbol{\Psi}_{2}$ by using the same method as calculating $\boldsymbol{\Psi}_{1}$ with other variables fixed;
		\STATE By fixing other variables, alternately calculate the auxiliary matrix $\boldsymbol{\Phi}_{1}$, $\boldsymbol{\Phi}_{2}$, $\mathbf{X}$ and $\mathbf{Y}$  in (\ref{phi_1}) and by the  bisection search method, respectively;
		\UNTIL{the difference in the values of the objective function between two consecutive iterations is less than the threshold $\delta$.
		}
	\STATE Update variables $\{\boldsymbol{\Gamma}_1,\boldsymbol{\Gamma}_2,\boldsymbol{\eta}_1,\boldsymbol{\eta}_2,\rho\}$ based on (\ref{dual_v_rho}).
	\UNTIL{the constraint violation indicator $P(\mathbb{X})$ is less than the threshold $\epsilon$ or the number of outer loop iteration is more than $t^{\textrm{max}}$.}
	\end{algorithmic}
\end{algorithm}

In the following, we analyze the complexity of the algorithm.
In (\ref{gamma}), the complexity of calculating $\boldsymbol{\gamma}$ is given by $\mathcal{O}(K(M_1^2+M_2^2))$. In (\ref{xi}), the complexity of calculating $\boldsymbol{\xi}$ is given by $\mathcal{O}(K(M_1^2+M_2^2))$. To solve Problem (\ref{C_W}), we calculate $\mathbf{W}(\boldsymbol{\lambda}^{(\tau)})$  in Algorithm \ref{AO_W_M} per iteration, where the orders of calculation complexity is $\mathcal{O}(N^3+KN^2)$, the computational complexity of ${\mathbf{g}}^{(\tau)}$ is given by $\mathcal{O}(KN^2+K^2N)$. Due to the fact that the dimension of Lagrange multipliers in Problem (\ref{C_W}) is much smaller than $M_1$ and $M_2$, the complexity of updating $\hat{\mathbf{g}}^{(\tau)}$, $\boldsymbol{\lambda}^{(\tau)}$ and $\boldsymbol{\Pi}^{(\tau)}$ can be neglected. Thus, the complexity of calculating the beamforming matrix $\mathbf{W}$ is approximately $\mathcal{O}(\tau^{\textrm{max}}(N^3+KN^2+K^2N))$. 
Similarly, the major computational complexity of calculating $\boldsymbol{\psi}_1$, $\boldsymbol{\psi}_2$ are given by $\mathcal{O}(\tau^{\textrm{max}}M_1^3)$ and $\mathcal{O}(\tau^{\textrm{max}}M_2^3)$, respectively. The major computational complexity of calculating $\boldsymbol{\Phi}_1$, $\boldsymbol{\Phi}_2$ can be ignored. The main computational complexity of calculating $\mathbf{X}$, $\mathbf{Y}$ are given by $\mathcal{O}(M_1^6)$ and $\mathcal{O}(M_2^6)$, respectively. In the general case $M_1\approx M_2 \gg N$, the approximate computational complexity of Algorithm \ref{PDD} is given by 
$\mathcal{O}(I_{o}I_{i}((M_1^6+M_2^6)+(\tau^{\textrm{max}}+K)(M_1^3+M_2^3)))$, where $I_o$, $I_i$ denote the maximum iteration numbers of the outer layer and inner layer of the PDD algorithm, respectively.

\section{Simulation Results}
In this section, simulation results are provided to evaluate the WSR performance of the proposed double-active-RIS-aided system and the efficiency of the proposed algorithm for the joint beamforming design.
\begin{figure}
	\centering
	\includegraphics[width=3.5in]{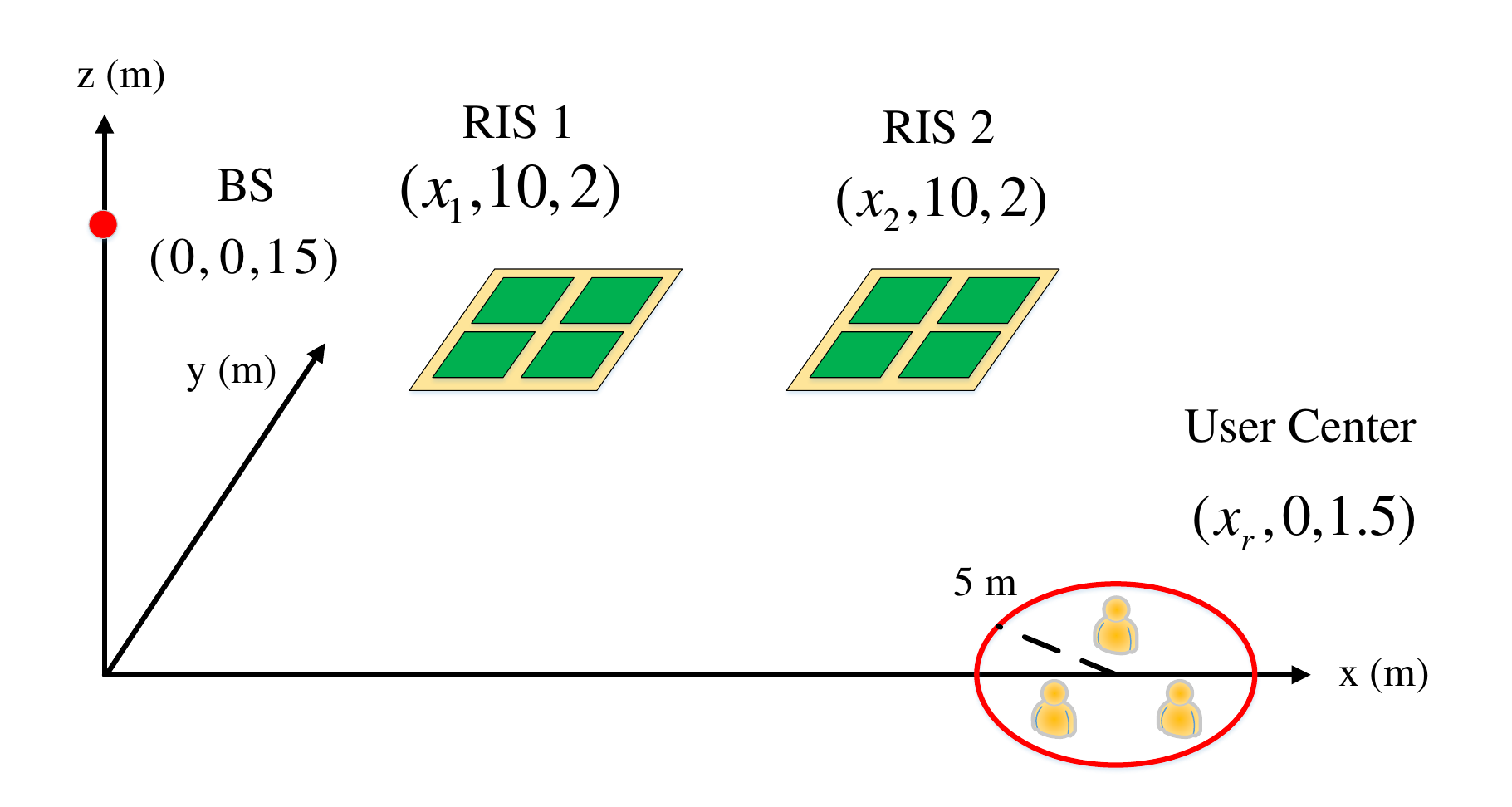}
	\caption{Simulation setup.}
	\label{Simulation_setup}
\end{figure}
We consider a three-dimensional scenario, where the BS is located at $(0 \ \textrm{m},0 \ \textrm{m},15 \ \textrm{m})$, the double active RISs are respectively deployed at $(x_1 \ \textrm{m},10 \ \textrm{m},2 \ \textrm{m})$ and $(x_2 \ \textrm{m},10 \ \textrm{m},2 \ \textrm{m})$. The users are uniformly distributed in a circle, where the center is $(x_u \ \textrm{m},0 \ \textrm{m},1.5 \ \textrm{m})$ and the radius is 5 m, as shown in Fig. \ref{Simulation_setup}.

Take ${\mathbf{H}_1}$ 
For example, ${\mathbf{H}_1}$ can be expressed as
\begin{align}
	{\mathbf{H}_1}=\sqrt{\beta_{\mathbf{H}_1} } \left ( \sqrt{\frac{\mathcal{F}_{\mathbf{H}_1}}{\mathcal{F}_{\mathbf{H}_1}+1} } {\mathbf{H}_{1,\textrm{LOS}}}+\sqrt{\frac{1}{\mathcal{F}_{\mathbf{H}_1}+1} }{\mathbf{H}_{1,\textrm{NLOS}}} \right ), 
\end{align}
where $\beta_{\mathbf{H}_1}$ denotes the large-scale path loss, $\mathcal{F}_{\mathbf{H}_1}$ is the Rician factor, ${\mathbf{H}_{1,\textrm{NLOS}}}$ represents the non-line-of-sight (NLOS) component of ${\mathbf{H}_1}$  following the Rayleigh distribution, and  ${\mathbf{H}_{1,\textrm{LOS}}}$ represents the LOS component of ${\mathbf{H}_1}$. We assume that the antennas/elements at the  BS and RISs are formed in a uniform linear array (ULA). Thus, ${\mathbf{H}_{1,\textrm{LOS}}}$ can be expressed as
\begin{align}
	{\mathbf{H}_{1,\textrm{LOS}}}=\mathbf{a}_r\left ( \theta^r \right ) \mathbf{a}_t^H\left ( \theta^t \right ), 
\end{align}
where the steering vectors of the BS and RIS 1 are respectively defined as
\begin{subequations}\label{steering}
\begin{align}
	\mathbf{a}_t\left ( \theta^t \right )&=\left [1,e^{j\frac{2\pi d_t}{\lambda }\textrm{sin} \theta^t  },\cdots, e^{j\frac{2\pi d_t}{\lambda }(N-1)\textrm{sin} \theta^t  } \right ]^T, \\
	\mathbf{a}_r\left ( \theta^r \right )&=\left [1,e^{j\frac{2\pi d_r}{\lambda }\textrm{sin} \theta^r  },\cdots, e^{j\frac{2\pi d_r}{\lambda }(M_1-1)\textrm{sin} \theta^r  } \right ]^T.
\end{align}
\end{subequations}
In (\ref{steering}), $\lambda$ is the wavelength of electromagnetic wave; $d_t$ and $d_r$ denote the antenna spacing distance at the BS and the element intervals of RIS 1; $\theta^t$ and $\theta^r$ are the angle of departure (AoD) from the BS and the angle of arrival (AoA) at RIS 1, respectively. We can generate other channels in the same process.

The large-scale path loss in dB can be modeled as
$
\textrm{PL}=\textrm{PL}_0-10\beta\textrm{log}_{10}(d/d_0 ),
$
where $\textrm{PL}_0=-30\ \textrm{dB}$ represents the large-scale path loss at the reference distance $d_0=1\ \textrm{m}$, $d$ represents the link distance in meters and $\beta$ represents the path loss exponent. 
In the simulation, we set the link path loss exponent to 2.2 for the links from BS to RIS 1, RIS 1 to RIS 2, and RIS 2 to the users, and 3.0 for the other link paths. This distinction is based on the fact that an increase in distance raises the likelihood of encountering additional obstacles and scatterers, which can result in significant signal attenuation.
%
The Rician factor is set to 10 for the links from the BS to RIS 1, from RIS 1 to RIS 2, and from RIS 2 to users, while the Rician factor is set to 1 for other links.
We set the space distance between the antennas at the BS and the reflecting elements of the active RISs as $d_{\textrm{BS}}=d_{\textrm{RIS}}=\lambda/2$, and we assume that the operating frequency of the proposed system is 2.4 $\textrm{GHz}$.

To demonstrate the effectiveness of the proposed beamforming design for the double-active-RIS-aided MU-MISO system with inter-excitation in improving the WSR performance, we conduct comparisons across the following scenarios:
\begin{itemize}
       \item[1)] \textbf{DAR-IE-Design (Double active RIS with beamforming design with inter-excitation effect)}:  In this scenario, the proposed Algorithm \ref{PDD} is applied to the beamforming design for the double-active-RIS-aided MU-MISO system with inter-excitation effect.
	\item[2)] \textbf{DAR-ideal (Double Active RIS without inter-excitation)}: In this scenario, we consider the double-active-RIS-aided MU-MISO system without the inter-excitation effect, which means that signals can only be reflected from RIS 1 to RIS 2. We set $\mathbf{X}=\mathbf{I}_{M_1}$, $\mathbf{Y}=\mathbf{I}_{M_2}$, remove the constraints (\ref{Xi_1_penalty}) and (\ref{Xi_2_penalty}), and skip the step of updating the auxiliary matrices $\mathbf{X}$ and $\mathbf{Y}$ in Algorithm \ref{PDD}. Then, the WSR performance is evaluated without the presence of inter-excitation effect. 
	\item[3)] \textbf{DAR-non-IE-Design (Double active RIS with beamforming design without considering inter-excitation effect)}: In this scenario, we use the same beamforming design as in “DAR-ideal”. Then, the WSR performance is evaluated under the presence of inter-excitation effect. Since the calculated $\boldsymbol{\Psi}_{1}$ and $\boldsymbol{\Psi}_{2}$ using the beamforming design in DAR-ideal may violate the constraint, we introduce a scaling factor $\tau \ (0<\tau\le1)$ to ensure that the final optimization results $\boldsymbol{\Psi}_{1}^{\textrm{opt}}=\tau\boldsymbol{\Psi}_{1}$ and $\boldsymbol{\Psi}_{2}^{\textrm{opt}}=\tau\boldsymbol{\Psi}_{2}$ satisfy the constraints (\ref{P_RIS1_E}) and (\ref{P_RIS2_E}).
	\item[4)] \textbf{SAR (Single active RIS)}: In this scenario, we consider the single-active-RIS-aided MU-MISO system. Denote $x_{A}$ as the horizontal coordinate of the single active RIS.
	 When the single active RIS is deployed near the BS, i.e., $x_{A}<x_u/2$, we set $\mathbf{H}_2=\mathbf{0}$, $\mathbf{G}=\mathbf{0}$,  $\mathbf{h}_{2,k}^H=\mathbf{0}$, $\mathbf{X}=\mathbf{I}_{M_1}$, $\mathbf{Y}=\mathbf{I}_{M_2}$.
	In this case, we skip steps 2 and 10, as well as the optimization of  auxiliary matrices $\{\boldsymbol{\Psi}_{2}, \mathbf{X}, \mathbf{Y}\}$ of 11 in the Algorithm \ref{PDD}. Conversely, when the single active RIS is deployed near the users, i.e., $x_{A}>x_u/2$, we set $\mathbf{H}_1=\mathbf{0}$, $\mathbf{G}=\mathbf{0}$, $\mathbf{h}_{1,k}^H=\mathbf{0}$, $\mathbf{X}=\mathbf{I}_{M_1}$, $\mathbf{Y}=\mathbf{I}_{M_2}$. In this case, we skip steps 2 and 9 as well as the optimization of auxiliary matrices $\{\boldsymbol{\Psi}_{1}, \mathbf{X}, \mathbf{Y}\}$ of 11 in the Algorithm \ref{PDD}.
%
	\item[5)] \textbf{DPR (Double passive RIS)}: In this scenario, we consider the double-passive-RIS-aided MU-MISO system. We set $\mathbf{X}=\mathbf{I}_{M_1}$, $\mathbf{Y}=\mathbf{I}_{M_2}$, $a^2_{\textrm{max}}=1$, $\sigma_{1}^2=\sigma_{2}^2=0$, relax the constraints (\ref{Xi_1_penalty}) and (\ref{Xi_2_penalty}) and skip the step of updating the auxiliary matrices $\mathbf{X}$ and $\mathbf{Y}$. 
\end{itemize}

It is assume that the total power $P_{\textrm{total}}$ and the number of the elements $M$ are the same for all benchmarks to make a fair comparison. Define $P_{\textrm{DC}}$ as the direct-current (DC) biasing power in each element of active RISs and $P_{\textrm{C}}$ as the power consumed by the switch and control circuit in each element of RISs. The hardware power consumption of the passive RISs can be denoted by $P_{\textrm{total}}^{\textrm{Pas}}=MP_{\textrm{C}}$, while the hardware power consumption of the active RISs can be denoted by $P_{\textrm{total}}^{\textrm{Act}}= M(P_{\textrm{C}}+P_{\textrm{DC}})$, respectively \cite{9734027}. 
Hence,  the total power consumption of the proposed system $P_{\textrm{total}}^{\textrm{DAR}}$ is given by 
\begin{align}
P_{\textrm{total}}^{\textrm{DAR}}&=P_{\textrm{BS}}^{\textrm{max}}+P_{1}^{\textrm{max}}+P_{2}^{\textrm{max}}+M(P_{\textrm{C}}+P_{\textrm{DC}}), 
\end{align}
and the total power
consumption models for “SAR” and “DPR” are respectively given as follows:
\begin{align}
P_{\textrm{total}}^{\textrm{SAR}}&=P_{\textrm{BS}}^{\textrm{max}}+P_{\textrm{A}}^{\textrm{max}}+M(P_{\textrm{C}}+P_{\textrm{DC}}),\\
P_{\textrm{total}}^{\textrm{DPR}}&=P_{\textrm{BS}}^{\textrm{max}}+MP_{\textrm{C}},
\end{align}
where we set $P_{\textrm{A}}^{\textrm{max}}=P_{1}^{\textrm{max}}+P_{2}^{\textrm{max}}$ to the transmit power allocated to the single active RIS.

Unless otherwise stated, the parameters are set as follows: the number of the antennas at the BS of $N=4$, the number of the elements at the active RISs of   $M_1=M_2=16$, the number of the users of $K=4$, the horizontal coordinate of active RIS 1 and active RIS 2 of $x_1=20 \ \textrm{m}$ and $x_2=40 \ \textrm{m}$, the horizontal coordinates of the center of the user of $x_u=60 \ \textrm{m}$, the weight of the users of $\alpha_k=1, \forall k\in \mathcal{K}$, the total power of the system of $P_{\textrm{total}}^{\textrm{DAR}}=30 \ \textrm{dBm}$, the transmit power of the active RISs of $P_{1}^{\textrm{max}}=P_{2}^{\textrm{max}}=14 \ \textrm{dBm}$, the DC biasing power of each element of active RISs of $P_{\textrm{DC}}=-5 \ \textrm{dBm}$, the power consumed by the switch and control circuit in each element of RISs of $P_{\textrm{C}}=-10 \ \textrm{dBm}$, the maximum amplification gain of the amplifier of $a_{\textrm{max}}^2=40 \ \textrm{dB}$\cite{9377648,10134546}, noise power of $\sigma_{1}^2=\sigma_{2}^2=\sigma^2=-80 \ \textrm{dBm}$, the Rician factor $\mathcal{F}_{\mathbf{H}_1}=\mathcal{F}_{\mathbf{H}_2}=\mathcal{F}_{\mathbf{G}}=\mathcal{F}_{\mathbf{h}_{1,k}}=\mathcal{F}_{\mathbf{h}_{2,k}}=10$.
\subsection{Convergence Behavior of the Proposed Algorithm}
\begin{figure}
	\centering
	\includegraphics[width=3.3in]{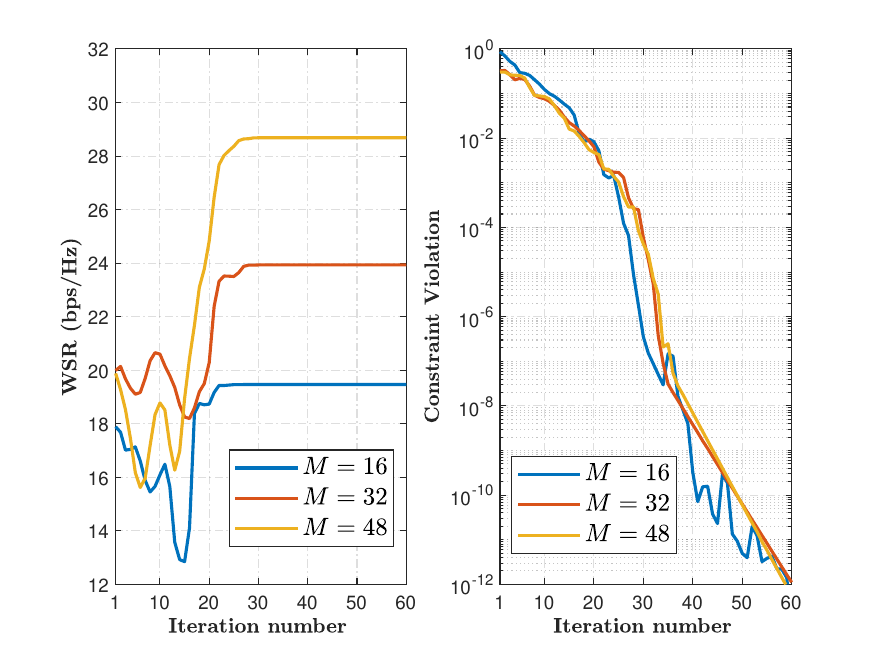}
	\caption{The convergence behavior of the Algorithm \ref{PDD}.}
	\label{Plot_Convergence_Speed}
\end{figure}

In Fig. \ref{Plot_Convergence_Speed}, we show the convergence behavior of the proposed PDD algorithm.
The WSR and the constraint violation indicator are provided versus the iteration number of the outer loop of the PDD framework while varying the total number of active RISs' reflecting elements, i.e., $M=16,32$ and 48 ($M_1=M_2=M/2$). 
In Fig. \ref{Plot_Convergence_Speed}(b), the constraint violation indicator $P(\mathbb{X})$ reduces to $10^{-10}$ within about 60 iterations of the outer loop of the PDD framework, which indicates
that the equality constraints  (\ref{penalty_constraint}) in Problem (\ref{PF_E}) are gradually satisfied.
It can be seen that the WSR does not increase monotonically with the outer iteration number of Algorithm \ref{PDD} for about the first 20 iterations, but rather fluctuates during the iteration procedure in Fig. \ref{Plot_Convergence_Speed}(a). 
It is due to the fact that the relatively large value of the penalty parameter $\rho$ leads to the severe violation to the equation constraints  (\ref{penalty_constraint}), which causes the oscillations of the value of the WSR. 
However, as the number of iterations increases, $\rho$ gradually approaches 0, forcing $P(\mathbb{X})$ to decrease gradually.
When $P(\mathbb{X})$ reaches an acceptable range ($P(\mathbb{X})\le 10^{-10}$), the WSR tends to stabilize.
Thus, Algorithm \ref{PDD} is ensured to converge eventually.

\subsection{Impact of the Maximum Amplification Gain}
\begin{figure}
	\centering
	\includegraphics[width=3.3in]{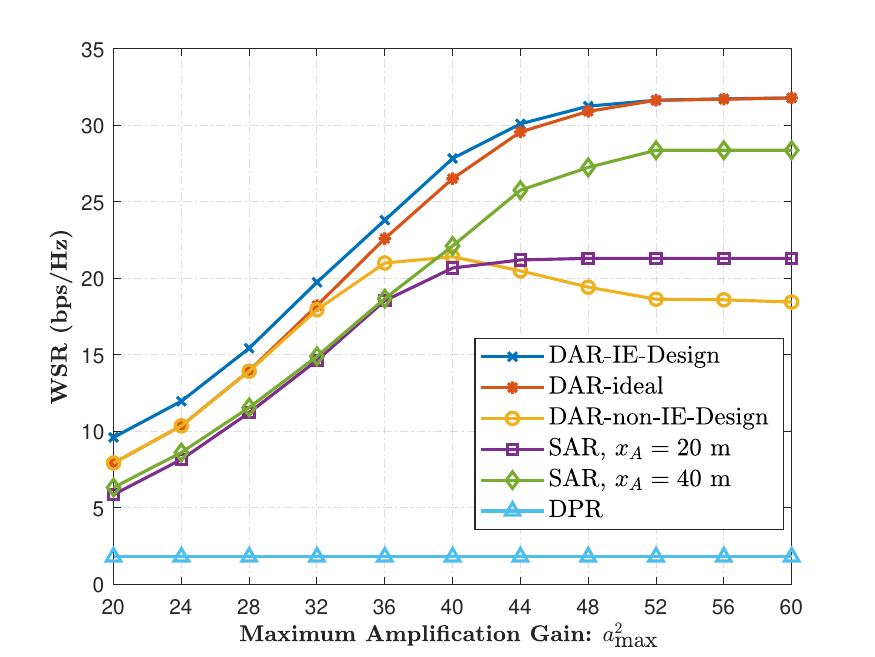}
	\caption{Achievable WSR versus the maximum amplification gain of the active RISs $a_{\textrm{max}}^2$. }
	\label{Plot_VS_a_max_2}
\end{figure}
In Fig. \ref{Plot_VS_a_max_2}, we show the WSR performance of various scenarios versus the maximum amplification gain of the active RISs.
It can be observed that the WSRs of “DAR-IE-design” are much higher than that of “SAR” and “DPR”. The reasons can be explained as follows: when a single reflecting link entails worse channel pathloss, i.e., when  $\beta=3$, providing an additional double reflecting link can help improve the WSR of the users. However, this performance improvement can only be achieved when the RISs are equipped with reflection-type amplifiers to compensate for the multiplicative fading effect.
As shown in Fig. \ref{Plot_VS_a_max_2}, the WSRs of “DAR-IE-Design”,  “DAR-ideal” and  “SAR” increase with the maximum amplification gain $a_{\textrm{max}}^2$, and then gradually flattens out with further increases $a_{\textrm{max}}^2$, i.e., $a_{\textrm{max}}^2\ge 52$ dB.
It indicates that when $a_{\textrm{max}}^2$ is relatively small, the transmit power of active RIS $P_l$ is limited by 
$a_{\textrm{max}}^2$, increasing $a_{\textrm{max}}^2$ can enhance $P_l$ and thereby improve the WSR performance. However, when 
$a_{\textrm{max}}^2$ is large, $P_l$ is constrained by the maximum transmission power of the active RIS, so increasing  $a_{\textrm{max}}^2$ does not improve the WSR performance.

It is worth noting that the WSR of “DAR-IE-Design”  is larger than the WSR of “DAR-ideal”, when $a_{\textrm{max}}^2 \le$ 52 dB. 
This means the inter-excitation effect can enhance the performance of the WSR when $P_l$ is constrained by $a_{\textrm{max}}^2$.
This is due to the fact that, in the presence of inter-excitation effect, the reflected signal from the other RIS contribute to the incident signal at the RIS
, thereby increasing $P_l$ and improving WSR performance.
Additionally, the disparity in the WSR between “DAR-IE-Design” and “DAR-ideal” diminishes with increasing $a_{\textrm{max}}^2$.
This shows that the gain from the inter-excitation effect decreases as $a_{\textrm{max}}^2$ increases. 
This is due to the fact that $P_l$ is progressively limited by $P_{l}^{\textrm{max}}$ as $a_{\textrm{max}}^2$ increases, and the inter-excitation effect is unable to raise the power of the reflected signals from the RIS exceed $P_{l}^{\textrm{max}}$.
Furthermore, it can be observed that when $a_{\textrm{max}}^2$ is less than 36 dB, “DAR-ideal” and “DAR-non-IE-Design” can reach almost the same WSR.
This finding reveals that the inter-excitation effect has a negligible effect on the WSR performance when $a_{\textrm{max}}^2$ is small.
However, as the maximum amplification gain increases, the disparity between the WSR of “DAR-non-IE-Design” and “DAR-IE-Design” grows much larger.
This shows that the influence of the inter-excitation effect on the system gradually increases as $a_{\textrm{max}}^2$ increases. Moreover, if the inter-excitation effect is not taken into account in the beamforming design it will result in a significant WSR performance loss.
Thus, it is crucial to consider the inter-excitation effect between active RISs in beamforming design.
\subsection{Impact of the Total Number of Reflecting Elements}
\begin{figure}
	\centering
	\includegraphics[width=3.3in]{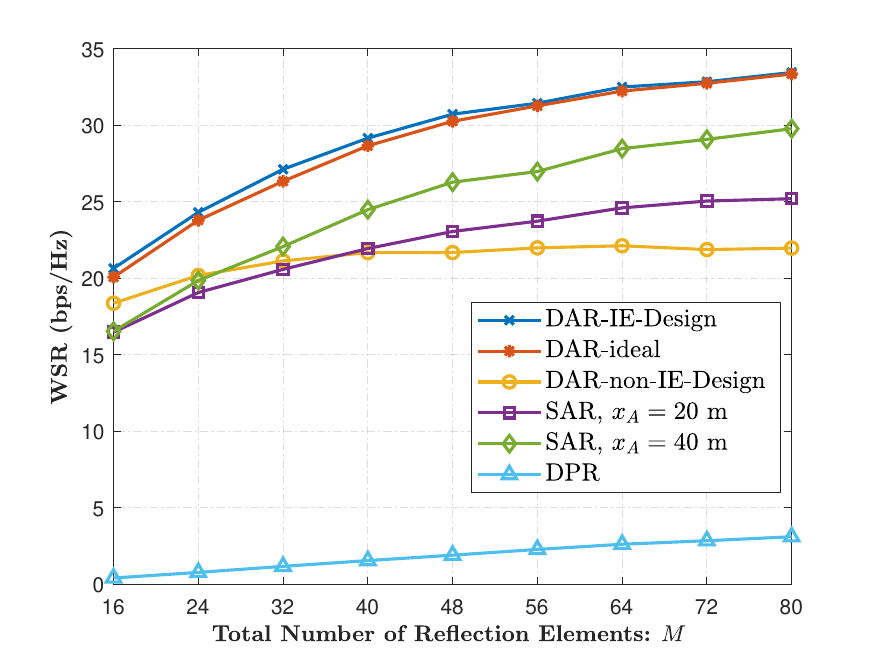}
	\caption{Achievable WSR versus the total number of reflecting elements $M$. }
	\label{Plot_VS_M}
\end{figure}
In Fig. \ref{Plot_VS_M}, we compare the WSR versus the total number of RIS elements for all scenarios. 
As shown in Fig. \ref{Plot_VS_M}, the WSR performance of all scenarios increases with the total number of reflecting elements $M$.
The WSR of the double-active-RIS-aided system with proper beamforming design significantly outperforms the single-active-RIS-aided system and the double-passive-RIS-aided system.
It reveals that the multiplicative fading effect can be better mitigated by the double-active-RIS-aided system.
Additionally, it is worth mention that the WSR performance of “DAR-non-IE-design” is even worse than that of “SAR” when $M$ is larger than 40.

Furthermore, it can be seen that the WSR performance gap between “DAR-non-IE-design” and “DAR-IE-design” increases with  $M$.
It indicates that as $M$ increases, the impact of inter-excitation effect on WSR performance becomes more pronounced.
Moreover, it can be observed that the WSR in “DAR-IE-Design” is slightly larger than the WSR in “DAR-ideal” when $M$ is relatively small. This is attributed to the fact that $P_l$ is limited by $a_{\textrm{max}}^2$ when $M$ is small.

\subsection{Impact of the Distance between the Active RISs}
\begin{figure}
	\centering
	\includegraphics[width=3.3in]{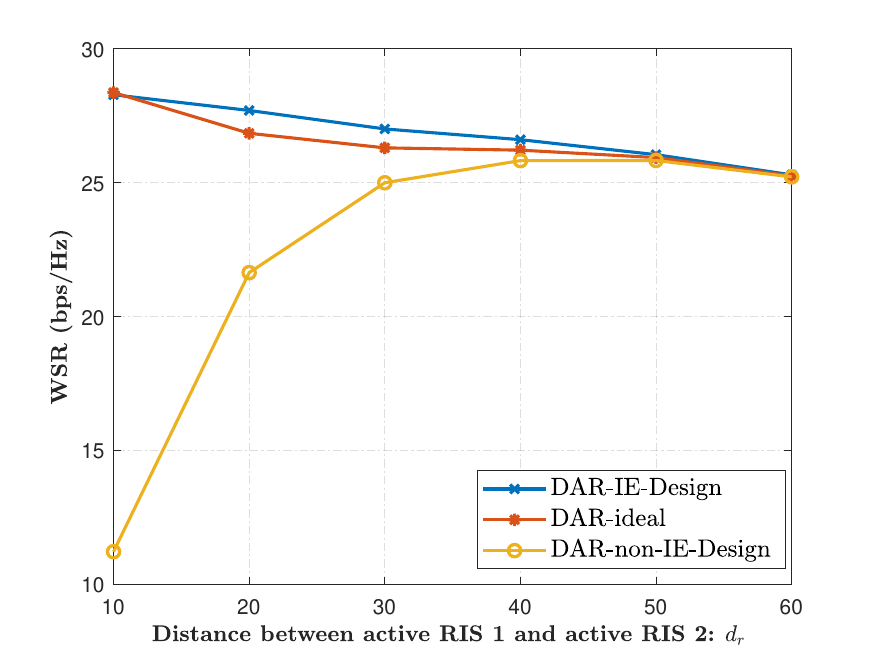}
	\caption{Achievable WSR versus the distance between the active RISs $d_r$. }
	\label{Plot_VS_Location}
\end{figure}
In Fig. \ref{Plot_VS_Location}, we examine the WSR versus the distance between the active RISs $d_r$ ranging from 10 m to 60 m.
The horizontal coordinates $x_1$ and $x_2$ of RIS 1 and RIS 2 are set to $x_1 = 30 - d_r/2$, $x_2 = d_r/2 + 30$.
It can be observed that the WSR of “DAR-IE-Design” and “DAR-ideal” gradually decreases with the increase of $d_r$.
It indicates that as $d_r$ increases, the transmitted power of the active RIS 2 decreases due to the constraints of $a_{\textrm{max}}^2$, which leads to a reduction in WSR.
Subsequently, it can be observed that the proposed scheme maintains high WSR across the entire distance range, unlike existing literature, which shows that double RIS configurations only perform optimally when placed near the BS and user, with poor performance in between (e.g., DAR-non-IE-Design). By considering the inter-excitation effect, the proposed scheme offers greater flexibility in positioning double RISs, overcoming the limitations of previous approaches.
Moreover, it can be observed that the performance gap between “DAR-non-IE-design” and “DAR-IE-design” diminishes with increasing $d_r$, and when $d_r$ exceeds 50, both designs achieve comparable WSR.
It indicates that the inter-excitation effect has a decreasing impact on the WSR of the double-active-RIS-aided system with the increase of the distance between the active RISs.
Therefore,  the “DAR-non-IE-design”  can be implemented when two active RISs are deployed at a considerable distance in practical systems. Although the WSR of this design is lower than that of the “DAR-IE-design” when the active RISs are positioned nearby, the performance loss remains within acceptable bounds, while the complexity of the beamforming design is greatly reduced.
This can be regarded as a trade-off between WSR performance and design complexity.

\subsection{Impact of the Allocation of Reflecting Elements}
\begin{figure}
	\centering
	\includegraphics[width=3.3in]{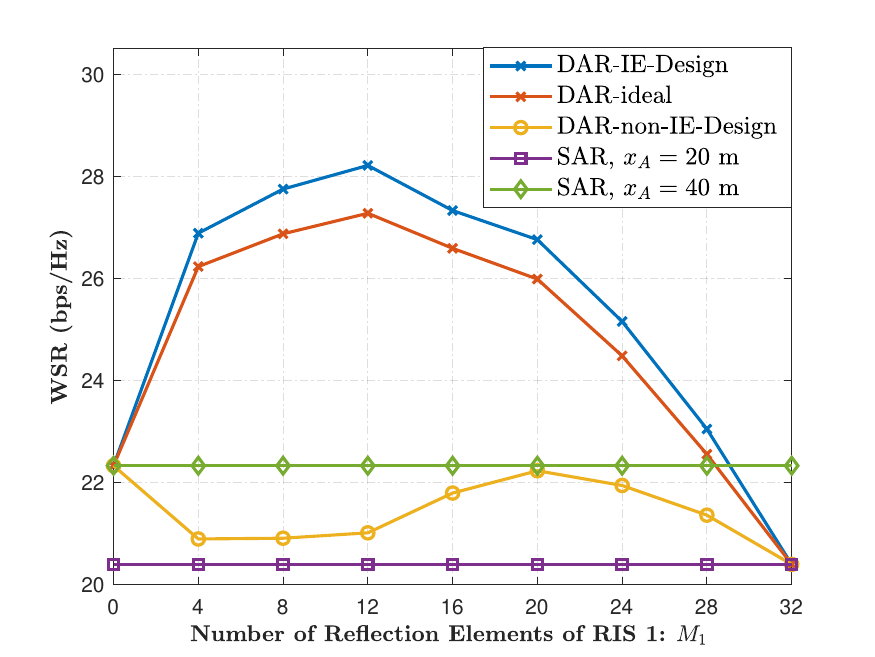}
	\caption{Achievable WSR versus the number of reflecting elements of RIS 1 $M_1$ given fixed total number of reflecting elements, i.e., $M=M_1+M_2=32$. }
	\label{Plot_VS_M_Allocation}
\end{figure}
In this subsection, we investigate the impact of reflecting element allocation in the double-active-RIS-aided system. In Fig. \ref{Plot_VS_M_Allocation}, we depicts the WSR versus the number of reflecting elements of RIS 1 $M_1$ while fixing the total number of the reflecting elements, i.e., $M=M_1+M_2=32$.
It can be observed that the WSR of the “DAR-IE-design” is higher than that of the “SAR” for most of reflection element allocation schemes when  the proposed beamforming design is employed. The improvement in WSR suggests that the incorporation of an additional double reflecting link can effectively mitigate the multiplicative fading effect. However, the WSR of the double-active-RIS-aided system that does not properly addressed the inter-excitation effect is lower than that of the single-active-RIS-aided system when the RIS is deployed near the users. 

In addition, it can be observed that the WSR performance gap between “DAR-non-IE-design” and “DAR-IE-design”, as well as the gap between “DAR-ideal” and “DAR-IE-design”,  initially increases and then decreases with increasing $M_1$.
This suggests that the impact of inter-excitation effect on WSR performance initially increases and then decreases as $M_1$ rises.
When $M_l$ is relatively small, the effective transmission power of the active RIS $P_l$ is constrained by the maximum amplification gain $a_{\textrm{max}}^2$. In this case, increasing $M_l$ can not only enhance the aperture gain but also increase the $P_l$. Conversely, when $M_l$ is relatively large, $P_l$ is limited by the maximum power budget of the active RIS $P_{l}^{\textrm{max}}$ , so increasing $M_l$ only increases the aperture gain.
Thus, the trend of WSR increasing with $M_1$ when $M_1$ is less than 12 implies that  the increase in the $P_1$ has a more significant positive impact on WSR than the decrease in the $P_2$. Conversely, the opposite effect occurs when $M_1$ exceeds 12. 
Moreover, the WSR performance gap between "DAR-non-IE-design" and "DAR-IE-design" is maximized when $M_1$ is slightly smaller than $M_2$. This occurs because the RIS 1, being closer to the BS than the RIS 2, receives relatively more energy from the incident signal. Consequently, the power constraint of RIS 1 are more easy to satisfied with fewer elements.
Therefore, allocating a sufficiently large number of elements to RIS 2 can enhance WSR performance.


\subsection{Impact of Transmit Power Allocation Between the Double Active RISs}
\begin{figure}
	\centering
	\includegraphics[width=3.3in]{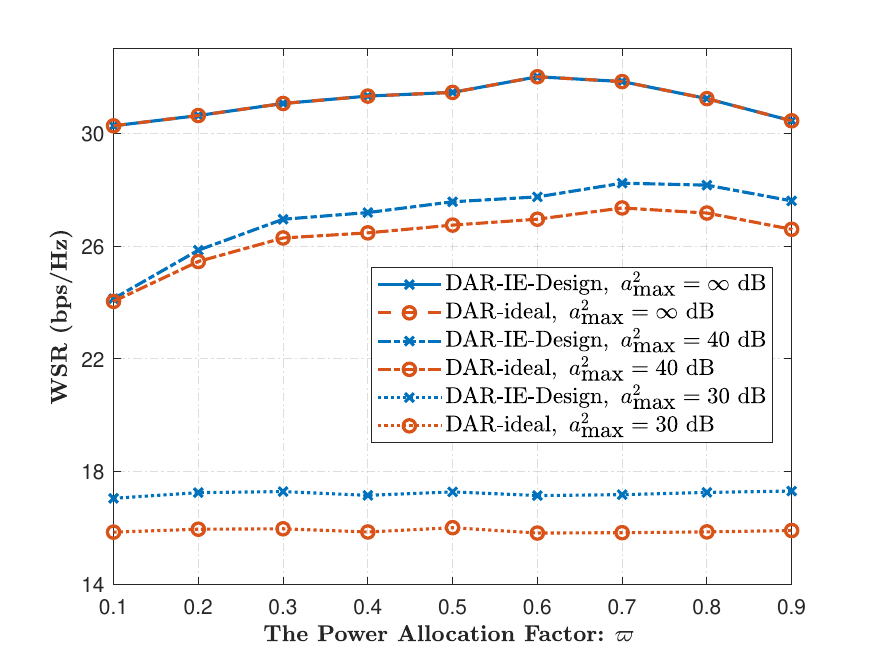}
	\caption{Achievable WSR versus the power allocation factor $\varpi$. }
	\label{Plot_VS_Power_Allocation}
\end{figure}
Denote $\varpi=P_{1}^{\textrm{max}}/(P_{1}^{\textrm{max}}+P_{2}^{\textrm{max}})$ as the power allocation ratio between two active RISs. In Fig. \ref{Plot_VS_Power_Allocation}, we concentrate on the impact of this power allocation on the WSR performance, considering various maximum amplification gains $a_{\textrm{max}}^2$  while maintaining the combined power of the two active RISs set to $P_{1}^{\textrm{max}}+P_{2}^{\textrm{max}}=17 \ \textrm{dBm}$.
It can be observed that appropriately allocating more power to RIS 1 can enhance the WSR performance when $a_{\textrm{max}}^2\ge 40 \textrm{dB}$.
Due to the fact that RIS 1 is closer to the BS than RIS 2, it receives relatively more energy from the incident signal. As a result, the effective transmitted power of RIS 1 is greater than that of RIS 2 with the same maximum amplification gain, leading to an improvement in the WSR.
It can also be observed that when $a_{\textrm{max}}^2= 30 \textrm{dB}$, changing the power allocation ratio has little effect on the WSR. This is because the transmission power of the RIS is constrained by $a_{\textrm{max}}^2$ under these conditions.

\section{Conclusions}
In this paper, we studied a double-active-RIS-aided MU-MISO system with inter-excitation effect. Specifically, we solved the WSR maximization problem by jointly optimizing the beamforming matrix at the BS and the reflecting coefficient matrices of the two active RISs, subject to power constraints at the BS and active RISs, as well as the maximum amplification gain constraints of the active RISs.
To address the non-convex objective function, we initially employed the FP method to transform the problem into a more tractable form.
Subsequently, we introduced auxiliary variables and equation constraints  to tackle the inter-excitation matrices.
Then, we applied the double-loop PDD framework to solve the introduced equational constraints.
Simulation results demonstrated that incorporating the inter-excitation effect into beamforming design for double-active-RIS-aided communication systems can enhance WSR performance, especially when the maximum amplification gain is substantial. 
Furthermore, the proposed scheme can achieve high WSR in most locations where the double active RIS is deployed between the BS and the users, thereby enhancing the flexibility in their positioning.

\bibliographystyle{IEEEtran}
\bibliography{ref}

\end{document}